\documentclass[reprint,twocolumn,superscriptaddress,showpacs,preprintnumbers,nofootinbib,aps,prd,floatfix]{revtex4-1}

\usepackage{graphicx}
\usepackage{amsmath,amssymb,amsfonts}
\usepackage{stackrel}
\usepackage{enumitem}
\usepackage[english]{babel}
\usepackage{verbatim}
\usepackage{comment}
\usepackage[normalem]{ulem}
\usepackage{xcolor}
\usepackage{dcolumn}
\usepackage{physics}
\usepackage{multirow}
\usepackage{makecell}
\usepackage{bm}
\usepackage{slashed}
\usepackage{float}
\usepackage{eso-pic}
\usepackage{placeins}

\definecolor{darkred}{rgb}{.7,.1,.1}
\definecolor{dark-green}{rgb}{0.1,0.7,0.3}

\newcommand{\tf}{\texorpdfstring}

\newcommand{\onbb}{0\nu\beta\beta}
\def\nn{\nonumber}

\graphicspath{{./}{figures/}}

\usepackage{hyperref}
\hypersetup{
    pdfnewwindow=true,
    colorlinks=true,       % NO BOXES/UNDERLINES
    linkcolor=darkred,
    citecolor=blue,        % CITATIONS = BLUE, NO UNDERLINE
    filecolor=blue,
    urlcolor=blue,
    pdfborder={0 0 0},     % REMOVE ALL BORDERS
    pdfencoding=unicode
}

\begin{document}

\title{Probing lepton flavor mixing in $W_R$ searches with machine learning at the LHC}% Force line breaks with \\

\author{Jin-Man Cai}
\email{caijm7@mail2.sysu.edu.cn}
\affiliation{School of Physics and Astronomy, Sun Yat-sen University, Zhuhai 519082, China}
\author{Gang~Li}%
\email{ligang65@mail.sysu.edu.cn}
\affiliation{School of Physics and Astronomy, Sun Yat-sen University, Zhuhai 519082, China}
\affiliation{
Guangdong Provincial Key Laboratory of Quantum Metrology and Sensing, Sun Yat-Sen University, Zhuhai 519082, China}

\date{\today}% It is always \today, today,

\begin{abstract}

Right-handed lepton flavor mixing in the left-right symmetric model directly affects the production and decay of heavy Majorana neutrinos $N_R$, yet its impact on collider searches remains less explored. Using a deep neural network (DNN), we analyze the Keung-Senjanović process $pp \to W_R \to \ell_\alpha N_R \to \ell_\alpha \ell_\beta jj$ with $\ell_{\alpha,\beta}=e,\mu$ at LHC Run~2 and the HL-LHC, considering both same-sign and opposite-sign dilepton channels. We adopt three benchmark mixing scenarios: unmixed, maximal-mixing, and PMNS-like. In the unmixed scenario, the DNN improves the expected significance over the cut-based analyses performed by ATLAS, leading to stronger expected sensitivities. For the combined $\ell\ell$ analysis, the HL-LHC is projected to reach $m_{W_R}$ and $m_{N_R}$ values up to $6.7$ ($6.3$)~TeV and $4.4$ ($4.1$)~TeV, respectively, under maximal (PMNS-like) mixing. For representative mass benchmarks, our analysis indicates that LHC Run~2 can test a significant portion of the $|V_{e1}|\text{--}|V_{\mu1}|$ plane, and the HL-LHC could probe even smaller mixing values, including the corresponding benchmark points for both the maximal and PMNS-like patterns. Finally, we investigate complementarities with low-energy charged lepton flavor violation processes, where future searches can overlap with or exceed the LHC reach.

\end{abstract}

\maketitle

%\tableofcontents

%%%%%%%%%%%%%%%%%%%%%%%%%%%%%%%%%%%%%%%%%
\section{Introduction}
\label{sec:intro}
%%%%%%%%%%%%%%%%%%%%%%%%%%%%%%%%%%%%%%%%%

The origin of neutrino masses remains one of the most compelling open questions in particle physics, motivating extensive searches for neutrino-related new physics beyond the Standard Model (SM). Among the proposed frameworks, the Left-Right Symmetric Model (LRSM)~\cite{Pati:1974yy,Mohapatra:1974gc,Senjanovic:1975rk,Senjanovic:1978ev} provides an especially attractive explanation for the smallness of neutrino masses. The model extends the SM gauge group to $SU(2)_L \times SU(2)_R \times U(1)_{B-L}$~\cite{Mohapatra:1979ia,Mohapatra:1980yp}, naturally introducing right-handed neutrinos. The spontaneous breaking of $SU(2)_R$ subsequently generates the heavy Majorana masses. Compared to the minimal type-I seesaw mechanism~\cite{Minkowski:1977sc,Gell-Mann:1979ijt,Gell-Mann:1979vob,Yanagida:1979as,Glashow:1979nm}, the LRSM establishes a more complete framework in which the origin of neutrino masses is directly connected to gauge symmetry breaking.

A key prediction of the LRSM is the existence of heavy gauge bosons $W_R^{\pm}$, $Z_R$ that couple with the SM fields and heavy Majorana neutrinos $N_R$.
At the Large Hadron Collider (LHC), the Keung-Senjanović process~\cite{Keung:1983uu} $pp \to W_R \to \ell N_R  \to \ell\ell jj$ via the mediation of $W_R$ serves as a smoking-gun signature for the LRSM.
This process not only enables direct searches for $W_R$ boson and the right-handed neutrino $N_R$, but also offers a direct way to probe the Majorana nature of $N_R$ through the observation of same-sign dilepton events.

Searches at the LHC during Run 2, with a center-of-mass energy $\sqrt{s} = 13~\text{TeV}$ and an integrated luminosity of $139~\text{fb}^{-1}$, have placed lower bounds on the masses of $W_R$ and $N_R$ at the TeV scale~\cite{ATLAS:2023cjo,CMS:2021dzb}. 
For instance, the ATLAS Collaboration excludes $m_{W_R}$ up to about $6.4$~TeV for $m_{N_R} = 1$~TeV~\cite{ATLAS:2023cjo}. However, these limits are obtained under the assumption of no lepton flavor mixing in the right-handed sector.
By contrast, the structure of right-handed neutrino mixing matrix $V_R$ in the LRSM has drawn increasing theoretical attention~\cite{Nemevsek:2012iq,Barry:2013xxa,BhupalDev:2014qbx,Senjanovic:2016vxw,Senjanovic:2018xtu,Senjanovic:2020int,Zhang:2020lir,Li:2024sln,Chen:2026jsc,Tello:2026ine}.

Notably, lepton flavor mixing effects in heavy-neutrino searches have been investigated experimentally in low-scale type-I seesaw models by both ATLAS and CMS at Run 2~\cite{CMS:2018jxx,CMS:2023jqi,ATLAS:2024rzi,CMS:2024ita,CMS:2024hik,ATLAS:2025uah}. Similar flavor-dependent searches in the context of the LRSM are lacking\,\footnote{In the type-I seesaw searches cited above, the flavor dependence is controlled by active--heavy neutrino mixing, rather than by the right-handed mixing matrix $V_R$. In the latter LRSM, an early ATLAS analysis at $\sqrt{s}=7~\text{TeV}$ with $2.1~\text{fb}^{-1}$ of data did consider lepton flavor mixing~\cite{ATLAS:2012ak}, but later searches omitted such possibilities~\cite{ATLAS:2015gtp,ATLAS:2018dcj,ATLAS:2019isd,ATLAS:2023cjo}.}.
In Ref.~\cite{Das:2012ii}, prospects for the $14~\text{TeV}$ LHC with an integrated luminosity of $30~\text{fb}^{-1}$ were investigated, including the effects of lepton flavor mixing. The determination of $V_R$ at the LHC was also discussed in Ref.~\cite{Vasquez:2014mxa} through ratios of flavor-dependent final states, with all $e$, $\mu$, and $\tau$ flavors taken into account.
Despite these phenomenological studies, a dedicated and standalone same-sign dilepton analysis was not performed, as it depends crucially on a proper treatment of SM backgrounds from fake and charge-misidentified leptons~\cite{ATLAS:2012ak}.
With the substantial improvement in experimental techniques and the much larger Run 2 dataset now available, it is timely and well motivated to derive updated 
flavor-dependent LHC constraints in the LRSM.

In this work, we conduct a detailed collider study of the Keung–Senjanović process at LHC Run 2 and the high-luminosity LHC (HL‑LHC) with $\sqrt{s}=14~\text{TeV}$ and $\mathcal{L}=3~\text{ab}^{-1}$, focusing on different 
$e,\mu$
flavor combinations in the final state. Motivated by the rapid development and widespread application of machine learning (ML) in high‑energy physics~\cite{Karagiorgi:2021ngt,Abdughani:2019wuv}, we perform an ML analysis of $W_R$ searches using a deep neural network~\cite{LeCun:2015pmr,Baldi:2014kfa,deOliveira:2015xxd}. Recent work~\cite{Chen:2025nzn} has also explored ML analyses of this process including lepton flavor mixing effects. 
Our work differs in several respects: beyond using a different ML method, we study the process within the minimal LRSM, while they considered the LRSM with inverse seesaw. Moreover, we include crucial SM backgrounds from fake and charge-misidentified leptons in the same‑sign dilepton channel, which are essential for a realistic collider analysis.

The rest of this paper is organized as follows. In Section~\ref{sec:model}, we briefly introduce the minimal LRSM.
Section~\ref{sec:KS-process} discusses the Keung-Senjanović process with possible flavor effects in the final state.
The detailed collider analyses, including background estimation and the ML analysis, are presented in Section~\ref{sec:collider_analysis}. In Section~\ref{sec:LHC_sensitivities}, we derive sensitivities to the masses of $W_R$ and $N_R$, as well as to the right-handed lepton mixing at LHC Run 2 and the HL-LHC.
We also study the complementarity with low-energy charged lepton flavor-violating (CLFV) processes in Section~\ref{sec:LHC_CLFV}. Finally, Section~\ref{sec:conclusion} summarizes the results.
More details of ML analysis and CLFV searches are provided in the appendices.

%%%%%%%%%%%%%%%%%%%%%%%%%%%%%%%%%%%%%%%%%
\section{Minimal Left-Right Symmetrical Model}
\label{sec:model}
%%%%%%%%%%%%%%%%%%%%%%%%%%%%%%%%%%%%%%%%%

The minimal LRSM~\cite{Pati:1974yy,Mohapatra:1974gc,Senjanovic:1975rk,Senjanovic:1978ev,Mohapatra:1979ia,Mohapatra:1980yp} extends the electroweak gauge symmetry of the SM to
$SU(2)_L \times SU(2)_R \times U(1)_{B-L}$. Each generation of quarks and leptons is assigned in doublets: 
\begin{align}
Q_{L,R} = \begin{pmatrix} u \\ d \end{pmatrix}_{L,R}\,, \quad 
L_{L,R} = \begin{pmatrix} \nu \\ \ell \end{pmatrix}_{L,R}\,.
\end{align}
The scalar sector includes a bidoublet and two triplet fields:
\begin{align}
\Phi &= \begin{pmatrix} \phi_1^0 & \phi_2^+ \\ \phi_1^- & \phi_2^0 \end{pmatrix}\,,\\
\quad \Delta_{L,R} &= \begin{pmatrix} {\Delta_{L,R}^+}/{\sqrt{2}} & \Delta_{L,R}^{++} \\ \Delta_{L,R}^0 & -{\Delta_{L,R}^+}/{\sqrt{2}} \end{pmatrix}\,.    
\end{align}
After the spontaneous symmetry breaking, they develop
vacuum expectation values as follows:
\begin{align}
    \langle\Phi\rangle &= \begin{pmatrix} \kappa/\sqrt{2} & 0 \\ 0 & \kappa^\prime e^{i\alpha}/\sqrt{2} \end{pmatrix}\,,\\
    \langle\Delta_{L}\rangle 
    &= \begin{pmatrix} 0 & 0 \\ v_Le^{i\theta_L}/\sqrt{2} & 0 \end{pmatrix}\,,\\
    \langle\Delta_{R}\rangle 
    &= \begin{pmatrix} 0 & 0 \\ v_R/\sqrt{2} & 0 \end{pmatrix}\,.
\end{align}
Here, $\alpha$ and $\theta_L$ are the spontaneous CP phases, which we assume to vanish for simplicity throughout this work. For a recent review of the model details, see Ref.~\cite{Kriewald:2024cgr}.

The weak eigenstates of charged gauge bosons $W_L$ and $W_R$ mix to form the mass eigenstates:
\begin{align}
\begin{pmatrix} W_1 \\ W_2 \end{pmatrix} = \begin{pmatrix} \cos\zeta & \sin\zeta \\ -\sin\zeta & \cos\zeta \end{pmatrix} \begin{pmatrix} W_L \\ W_R \end{pmatrix}\,.
\end{align}
The left-right mixing parameter $\zeta$ is defined as
\begin{align}
    \sin \zeta = \lambda \sin(2\beta)\,,
\end{align}
with 
$\tan\beta \equiv \kappa^\prime/\kappa$ and $\lambda \simeq m^2_{W_1}/m^2_{W_2}$.
The experimental constraints on $\lambda$ give $\lambda \lesssim 10^{-4}$~\cite{deVries:2022nyh,Li:2025tmt}. 
Thus, we can safely neglect the left-right mixing\,\footnote{For the nontrivial impact of $\zeta$ on low-energy observables, see Refs.~\cite{Li:2020flq,Ramsey-Musolf:2020ndm}.}, and treat $W_{L,R}$ as $W_{1,2}$, respectively.

The charged-current interactions are described as
\begin{align} 
\label{eq:lepton_CC}
\mathcal{L}^{\ell}_{CC} &= \frac{g}{\sqrt{2}} \sum_{\ell = e,\mu,\tau} \sum_{i=1}^{3} \Big[ \bar{\ell}_{L} \slashed{W}_{L}^{-} V_{L\ell i} \nu_{i} \notag \\
&\qquad + \bar{\ell}_{R} \slashed{W}_{R}^{-} V_{R\ell i} N_{i} \Big] + \text{h.c.}\,, \\
\label{eq:quark_CC}
\mathcal{L}^{q}_{CC} &= \frac{g}{\sqrt{2}} \sum_{i,j=1}^{3} \Big[ \bar{u}_{Li} \slashed{W}_{L}^{+} V^{q}_{Li j}d_{Lj} \notag \\
&\qquad + \bar{u}_{Ri} \slashed{W}_{R}^{+} V^{q}_{Rij} d_{Rj} \Big] + \text{h.c.}\,,
\end{align}
where we have assumed that the gauge couplings are equal, $g_L = g_R \equiv g \simeq 0.65$. 
In the above, $\nu_i$ and $N_i$ $(i=1,2,3)$ denote the light and heavy neutrinos, respectively. They are related to the left-handed neutrino $\nu_{L\alpha}$ and right-handed neutrino $\nu_{R\alpha}$ in the flavor basis $(\alpha=e,\mu,\tau)$ via lepton flavor mixing\,\footnote{We work in the charged-lepton mass basis, in which the charged-lepton mass matrices are taken to be diagonal~\cite{Das:2012ii}. This choice is without loss of generality in the minimal LRSM for both generalized parity~\cite{Senjanovic:2018xtu} and generalized charge conjugation~\cite{Nemevsek:2012iq}.
}:
\begin{align}
\label{eq:mass-flavor}
    \nu_{L \alpha} = V_{L\alpha i} \nu_i\,, \quad
    \nu_{R\alpha} = V_{R\alpha i} N_i\,,
\end{align}
where $V_L = V_{\rm PMNS}$ with  $V_{\rm PMNS}$ being the Pontecorvo-Maki-Nakagawa-Sakata (PMNS) matrix, and $V_R$ is the right-handed counterpart of $V_L$. 
$V^{q}_{L}$ denotes the Cabibbo-Kobayashi-Maskawa matrix describing flavor mixing in the quark sector, while $V^{q}_{R}$ is its right-handed analogue.
For generalized parity, $V^{q}_{R} \simeq V^{q}_{L}$ up to a small correction~\cite{Senjanovic:2014pva}, while for generalized charge conjugation the two matrices are equal up to an overall phase~\cite{Maiezza:2010ic}.

The leptonic Yukawa Lagrangian is also relevant, as it can lead to CLFV processes. The interactions are given by
\begin{align} \label{eq:yukawa}
-\mathcal{L}_Y^\ell= \bar L_L^c \epsilon \Delta_L f_L L_L + \bar L_R^c \epsilon \Delta_R f_R L_R + \text{h.c.}\,,
\end{align}
where $f_L$ and $f_R$ denote the general Yukawa coupling matrices, and $\epsilon \equiv i\tau^2$ with $\tau^2$ the second Pauli matrix.

%%%%%%%%%%%%%%%%%%%%%%%%%%%%%%%%%%%%%%%%%
\section{Flavored Keung-Senjanović process}
\label{sec:KS-process}
%%%%%%%%%%%%%%%%%%%%%%%%%%%%%%%%%%%%%%%%%

\begin{figure}[!htbp]
\centering
\includegraphics[width=0.4\textwidth]{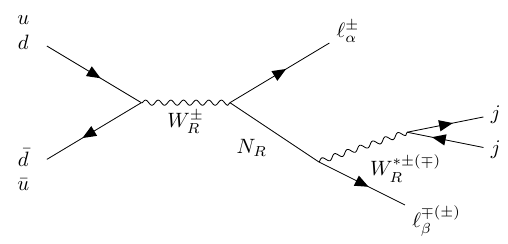}
\caption{Feynman diagram of the Keung-Senjanović process at the LHC, illustrating different flavor and electric charge combinations of the charged leptons.
}
\label{fig:KS_process} 
\end{figure}

At the LHC, $W_R$ boson can be produced via the charged-current interactions and decay into a charged lepton and a heavy neutrino. Owing to the Majorana nature of the heavy neutrino, it subsequently decays into a second charged lepton, which can carry either the same or opposite electric charge as the first. This process is known as the Keung-Senjanović process~\cite{Keung:1983uu}, as illustrated in Fig.~\ref{fig:KS_process}.

We assume that $N_1$ is the lightest heavy neutrino, which we denote as $N_R$, and that it is well separated in mass from the other two heavy neutrinos $N_{2,3}$ and lighter than $W_R$ boson. Thus, $N_R$ is taken to be on shell, with its production and decay factorized in the narrow-width approximation, and interference among different heavy neutrinos is neglected.
The parton-level cross section for the production of $N_R$ via $pp \to \ell_\alpha^\pm N_R$, where $\ell_\alpha$ denotes the charged lepton of flavor $\alpha$, is given by~\cite{Nemevsek:2018bbt}\,\footnote{As in Ref.~\cite{Nemevsek:2018bbt}, we do not apply the narrow-width approximation to the intermediate $W_R$, and retain the full contribution from $W_R$ exchange.}
\begin{align}
\label{cross:lN}
\hat{\sigma}_{ij}^{\ell N}(\hat{s})&=\frac{\alpha_W^2\pi\lvert V_{R\alpha1}\rvert ^2 \lvert V^{q}_{Ri j}\rvert^2}{72\hat{s}^2}\nn\\
& \quad \times \frac{\left(\hat{s}-m_{N_R}^2\right)^2\left(2\hat{s}+m_{N_R}^2\right)}{\left(\hat{s}-m_{W_R}^2\right)^2+m_{W_R}^2\Gamma_{W_R}^2},
\end{align}
where $\alpha_W = g^2/(4\pi)$, $\hat{s}$ denotes the squared partonic center-of-mass energy, and $\Gamma_{W_R}$ is the decay width of $W_R$.

The on-shell heavy neutrino $N_R$ decays via the three-body process $N_R \to \ell_\beta^{\pm} jj^\prime$, mediated by an off-shell $W_R$ boson. The corresponding decay width is~\cite{Helo:2013esa,Vasquez:2014mxa}
\begin{align}
\label{width:jjl}
\Gamma(N_R \to \ell^{\pm}_\beta jj^\prime ) =N_c \frac{g^4 \lvert V_{R \beta1}\rvert ^2}{768(2\pi)^3 } \frac{m_{N_R}^5}{m_{W_R}^4} \sum_{jj^\prime} |V^{q}_{Rjj^\prime}|^2\,,
\end{align}
where $N_c$ is the number of colors, and $j, j^\prime$ denote light quarks or anti-quarks\,\footnote{A complementary LHC search targeting the heavy-flavor topology
$pp\to W_R\to \ell N_R\to \ell\ell t b$ was proposed in
Ref.~\cite{Frank:2023epx}, together with a recast of the existing CMS search
at LHC Run 2 for the Keung-Senjanovi\'c process in the same-flavor channels~\cite{CMS:2021dzb}.}. From the expressions in Eqs.~\eqref{cross:lN} and~\eqref{width:jjl}, 
the signal depends on the first column of $V_R$.

The total decay width sums over all lepton flavors. From the unitarity condition
\begin{align}
\label{eq:unitary}
    \lvert V_{Re1} \rvert^2 + \lvert V_{R\mu 1} \rvert^2 + \lvert V_{R\tau 1} \rvert^2 = 1\,,
\end{align}
the total width does not depend on the individual mixing entries. Consequently, the event rate for $pp \to \ell_\alpha \ell_\beta j j^\prime $ is proportional to $|V_{R\alpha1}|^2 |V_{R\beta 1}|^2$.
Final states involving $\tau$ leptons are not included in our analysis due to the experimental challenges associated with $\tau$ reconstruction, and $V_{R\tau1}$ enters only indirectly through the unitarity condition in Eq.~\eqref{eq:unitary}. Thus, for the signal process, the flavor indices are restricted to $\alpha,\beta=e,\mu$.

Recent experimental searches at the LHC have placed stringent bounds on the masses of $W_R$ boson and heavy neutrino $N_R$. Using both opposite-sign (OS) and same-sign (SS) dilepton channels, the ATLAS Collaboration~\cite{ATLAS:2023cjo} excludes $m_{W_R} \lesssim 5\text{--}6~\text{TeV}$ for $m_{N_R} = 1\text{--}3~\text{TeV}$ in the resolved analysis. The OS dilepton channel achieves sensitivity comparable to that of the SS channel in this mass range, while the latter provides a distinctive signature of lepton number violation.

However, most existing studies, including the experimental analyses above, have focused exclusively on same-flavor charged leptons ($ee$ and $\mu\mu$), neglecting the possibility of lepton flavor violation induced by a nontrivial right-handed lepton mixing matrix $V_R$. In the presence of flavor mixing, charged leptons of different flavors ($e\mu$) also appear in the final state, and the signal rates depend sensitively on the flavor structure of $V_R$. Moreover, flavor effects significantly affect the experimental bounds on $m_{W_R}$ and $m_{N_R}$ even in the $ee$ and $\mu\mu$ channels, since they alter both the production cross section of $N_R$ and its decay branching ratios.

The expected numbers of SS dilepton events for the Keung-Senjanović process in the final states $e^\pm e^\pm jj$, $\mu^\pm \mu^\pm jj$, and $e^\pm \mu^\pm jj$ are given by
\begin{subequations}
\label{eq:events_SS}
\begin{align}
N(e^{\pm} e^{\pm}) &= \mathcal{L} \cdot \epsilon_{ee}^{\rm SS} \cdot \sigma^{\rm SS}_{ee} \cdot \lvert V_{e1} \rvert^4\,,\\
N(\mu^{\pm}\mu^{\pm}) &= \mathcal{L} \cdot \epsilon_{\mu\mu}^{\rm SS} \cdot \sigma^{\rm SS}_{\mu\mu} \cdot \lvert V_{\mu1} \rvert^4\,,\\
\label{events_emu}
N(e^{\pm}\mu^{\pm}) &= \mathcal{L} \cdot \epsilon_{e\mu}^{\rm SS} \cdot \sigma^{\rm SS}_{ee} \cdot 2 \cdot \lvert V_{e1} \rvert^2 \cdot \lvert V_{\mu1} \rvert^2\,.
\end{align}
\end{subequations}
Here, $\sigma_{ee}^{\rm SS}$ and $\sigma_{\mu\mu}^{\rm SS}$ are the inclusive cross sections, summed over the two charge configurations, under the assumptions $\lvert V_{e1} \rvert = 1$ and $\lvert V_{\mu1} \rvert = 1$, respectively, which are equal since the masses of $e$ and $\mu$ are negligible.
The quantities $\epsilon_{ee}^{\rm SS}$, $\epsilon_{\mu\mu}^{\rm SS}$, and $\epsilon_{e\mu}^{\rm SS}$ are the corresponding overall selection efficiencies, and $\mathcal{L}$ is the integrated luminosity. For convenience, we denote the right-handed lepton mixing matrix simply as $V$ (i.e., $V_R \to V$). In Eq.~\eqref{events_emu}, the factor of 2 accounts for the two possible flavor sequences in which the electron and muon can appear.

For expected numbers of signal events in the OS dilepton channels $e^+ e^-$, $\mu^+ \mu^-$, and $e^\pm \mu^\mp$, analogous expressions hold~\cite{Das:2012ii}:
\begin{subequations}
\label{eq:events_OS}
    \begin{align}
    N(e^{+} e^{-}) &= \mathcal{L} \cdot \epsilon_{ee}^{\rm OS} \cdot \sigma^{\rm OS}_{ee} \cdot \lvert V_{e1} \rvert^4\,,\\
    N(\mu^{+}\mu^{-}) &= \mathcal{L} \cdot \epsilon_{\mu\mu}^{\rm OS} \cdot \sigma^{\rm OS}_{\mu\mu} \cdot \lvert V_{\mu1} \rvert^4\,,\\
    \label{eq:events_emu_b}
    N(e^{\pm}\mu^{\mp}) &= \mathcal{L} \cdot \epsilon_{e\mu}^{\rm OS} \cdot \sigma^{\rm OS}_{ee} \cdot 2  \cdot \lvert V_{e1} \rvert^2 \cdot \lvert V_{\mu1} \rvert^2\,,
\end{align}
\end{subequations}
where the quantities carrying the superscript OS follow definitions similar to their SS counterparts.

%%%%%%%%%%%%%%%%%%%%%%%%%%%%%%%%%%%%%%%%%
\section{Collider analyses}
\label{sec:collider_analysis}
%%%%%%%%%%%%%%%%%%%%%%%%%%%%%%%%%%%%%%%%%

The \texttt{FeynRules}~\cite{Alloul:2013bka} implementation of the LRSM and the corresponding UFO model are publicly available~\cite{Kriewald:2024cgr,LRSM_NLO}. Signal and SM background events for both SS and OS channels are generated with \texttt{MadGraph5\_aMC@NLO}~\cite{Alwall:2014hca}, using the NNPDF3.0 PDF set via \texttt{LHAPDF6}~\cite{Buckley:2014ana}. Within this setup, the decay widths of both $N_R$ and $W_R$ are computed automatically.
The following {\it basic} cuts are applied at the generator level:
\begin{align}
p_{T_{j}}> 20~\text{GeV}\,,\quad \vert \eta_{j} \vert<2.8\,,\nn\\
p_{T_{\ell}}> 20~\text{GeV}\,,\quad \vert \eta_{\ell} \vert<2.5\,.
\label{eq:basic_cuts}
\end{align}
Here, $p_{T_{j}}$ and $p_{T_{\ell}}$ denote the transverse momenta of quarks and leptons, while $\eta_{j}$ and $\eta_{\ell}$ are the corresponding pseudorapidities. The parton-level events are passed to \texttt{Pythia8}~\cite{Sjostrand:2014zea} for hadronization and \texttt{Delphes3}~\cite{deFavereau:2013fsa} for fast detector simulation. In the SS (OS) dilepton channels, the {\it baseline} selection requires events to contain exactly two charged leptons with the same (opposite) electric charges and at least two jets.

% --------------------------------------
\subsection{Simulation of SM Backgrounds}
% --------------------------------------

In searches for the Keung‑Senjanović process, modeling of SM backgrounds is essential, particularly in the SS dilepton channel, where such backgrounds are rare. The backgrounds contributing to the SS dilepton final state fall into three categories:
\begin{itemize}
    \item Prompt backgrounds: arising from SM diboson processes ($W^\pm W^\pm$, $WZ$, $ZZ$).
    \item Fake-lepton backgrounds: originating from jets misreconstructed as leptons or photon conversions ($t\bar{t}$, $W$+jets). 
    \item Charge-flip (CF) backgrounds: arising from SM processes with OS leptons ($t\bar{t}$, $Z/\gamma^*$+jets, $W^+W^-$), where the electric charge of an electron ($e^\pm$) is misidentified.  
\end{itemize}
The dominant processes for each category are indicated in parentheses. These backgrounds have been considered in earlier experimental searches~\cite{ATLAS:2023cjo,CMS:2021dzb,CMS:2017tec,ATLAS:2017xqs,CMS:2018jxx} and phenomenological studies of TeV-scale lepton number violation at colliders~\cite{Peng:2015haa,Harz:2021psp}. 

In the OS dilepton channels, prompt backgrounds are dominant. Specifically, we account for backgrounds originating from diboson production ($W^+W^-$, $WZ$, $ZZ$), top pair production ($t\bar{t}$), and single boson production ($Z/\gamma^*$+jets)~\cite{ATLAS:2023cjo,Chen:2025nzn}. While prompt backgrounds are generated directly via Monte Carlo (MC) simulation, the other backgrounds are modeled as described below.

Instead of the data-driven methods employed in experimental analyses, Ref.~\cite{Curtin:2013zua} proposed a jet-fake approach to estimate fake-lepton backgrounds, which we refer to as jet-fake (JF) backgrounds. This method uses jet‑enriched samples to model the fake‑lepton backgrounds via two functions: the mistag efficiency and the transfer functions. The mistag efficiency function is defined as
\begin{equation}
\epsilon_{j \rightarrow \ell}(p_{T_j}) = \epsilon_{200} \left[ 1 - (1 - r_{10}) \frac{200 - p_{T_j}/\text{GeV}}{200 - 10} \right],
\end{equation}
where $\epsilon_{j \rightarrow \ell}(p_{T_j})$ 
denotes the probability that a jet with transverse momentum $p_{T_j}$ is mistagged as a prompt lepton. 
The transfer function, which maps the transverse momentum of the jet $p_{T_j}$ to the reconstructed transverse momentum of the fake lepton $p_{T_\ell}$, is parameterized by a truncated Gaussian distribution as follows:
\begin{equation}
T_{j \rightarrow \ell} = \frac{1}{N} \exp \left[ -\frac{(\alpha - \mu)^2}{2\sigma^2} \right]\,,
\end{equation}
where $N$ is the normalization factor, and the parameter $\alpha \equiv 1-p_{T_\ell}/p_{T_j}$ satisfies $0 < \alpha < 1$. 

This jet‑fake method has been widely used in the literature~\cite{Izaguirre:2015pga,Nemevsek:2016enw,Dib:2016wge,Dib:2017iva,Harz:2021psp}. In this work, we implement it as a dedicated module in \texttt{Delphes3}~\cite{Nemevsek:2016enw,Harz:2021psp} and estimate the dominant backgrounds $t\bar{t}$ and $W$+jets. The parameters $\{\epsilon_{200}, r_{10}, \mu, \sigma\}$ are then fitted by comparing the simulated differential distributions with those reported by the ATLAS or CMS Collaboration.

Due to bremsstrahlung effects, the electric charge of electrons can be misidentified. Consequently, SM backgrounds arising from charge misidentification of electrons must be taken into account in SS dilepton searches. In contrast, the charge misidentification rate for muons is negligible and is therefore ignored. Following the methodology in Refs.~\cite{ATLAS:2017xqs,Muskinja:2018eba}, the charge misidentification probability for electron, $P(p_T, \eta)$, is parameterized as
\begin{align}
    P(p_T, \eta) = \sigma(p_T) f(\eta)\,,
\end{align}
where $\sigma(p_T)$ and $f(\eta)$ encode the dependence on the electron transverse momentum and pseudorapidity, respectively. These functions are extracted from a likelihood fit to a dedicated $Z \to e^+ e^-$ data sample~\cite{ATLAS:2017xqs}\,\footnote{As in Ref.~\cite{Harz:2021psp}, we validate this modeling by comparing the ratio of SS to OS event yields from our simulated $Z\to e^+e^-$ events with that from the ATLAS results, and find overall good agreement, with a relative difference of about $17\%$. This provides an estimate of the associated systematic uncertainty, consistent
with the evaluation in Ref.~\cite{Muskinja:2018eba}, where deviations of up to
about $20\%$ in the charge-misidentification rate were found in some pseudorapidity
regions for the adopted parametrization.}.

To verify the overall background modeling, we apply the following selection criteria to events from all three background categories~\cite{ATLAS:2023cjo}:
\begin{align}
& p_{T_{\ell_1}} > 40~\text{GeV}\,,\quad m_{\ell\ell} > 400~\text{GeV}\nn\,,       \\
& \Delta R(\ell_1,\ell_2) < 3.9\,,\quad m_{jj} > 110~\text{GeV}\,,\nn\\
& p_{T_{j_1}} > 100~\text{GeV}\,,\quad p_{T_{j_2}} > 100~\text{GeV}\nn\,,         \\
& H_T \equiv p_{T_{j_1}}+p_{T_{j_2}}+p_{T_{\ell_1}}+p_{T_{\ell_2}} > 400~\text{GeV}\,.
\label{eq:atlas_selection}
\end{align}
Here, $p_{T_{\ell_1}}$ is the transverse momentum of the leading lepton, while $p_{T_{j_1}}$ and $p_{T_{j_2}}$ are the transverse momenta of the leading and subleading jets, respectively. Furthermore, $\Delta R(\ell_1,\ell_2)$ denotes the angular separation between the two leptons, $m_{\ell\ell}$ ($m_{jj}$) is the dilepton (dijet) invariant mass, and $H_T$ is the scalar sum of the transverse momenta of the two leading leptons and two leading jets.

\begin{figure*}
  \centering
  \includegraphics[width=0.32\textwidth]{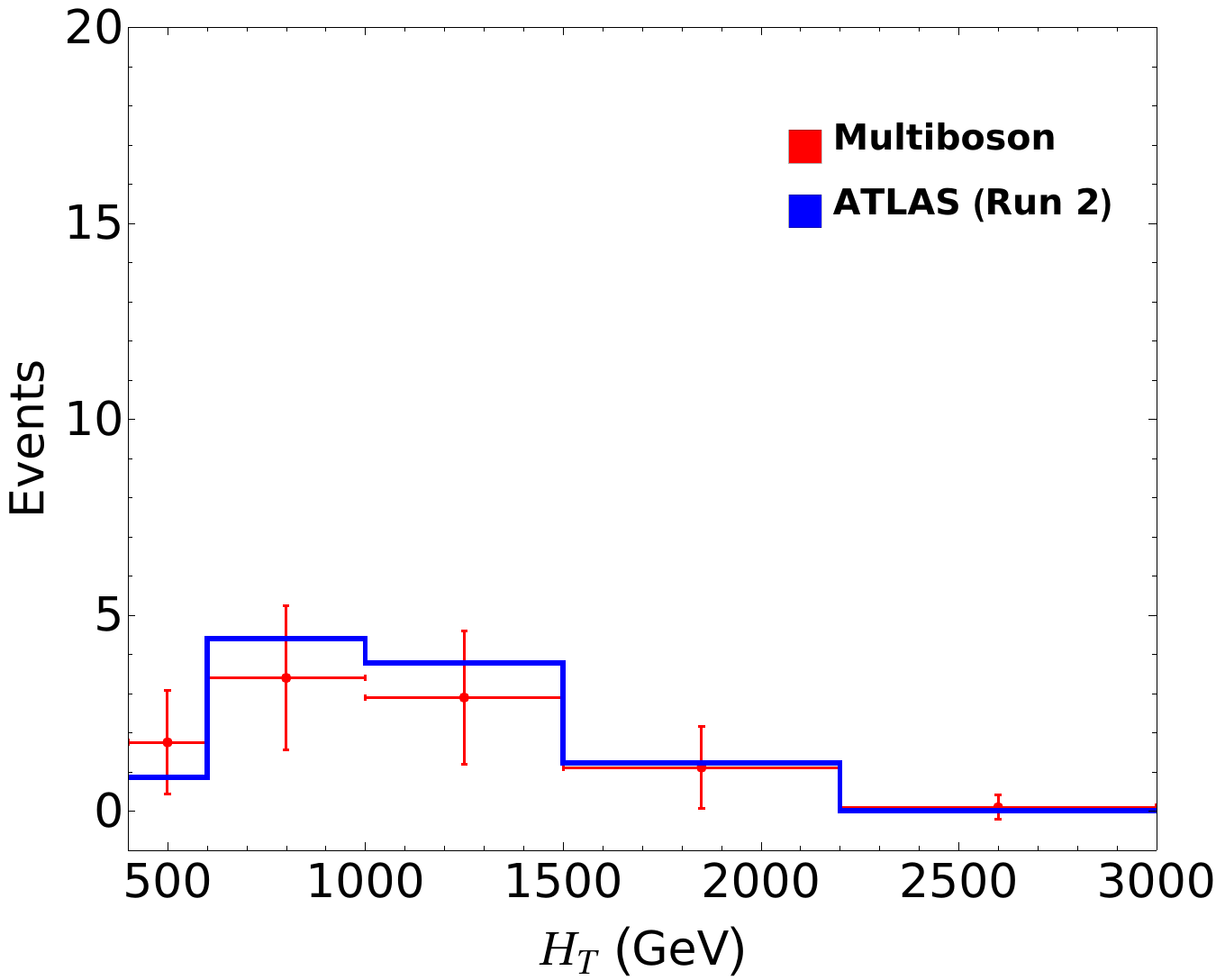}
  \hfill
  \includegraphics[width=0.32\textwidth]{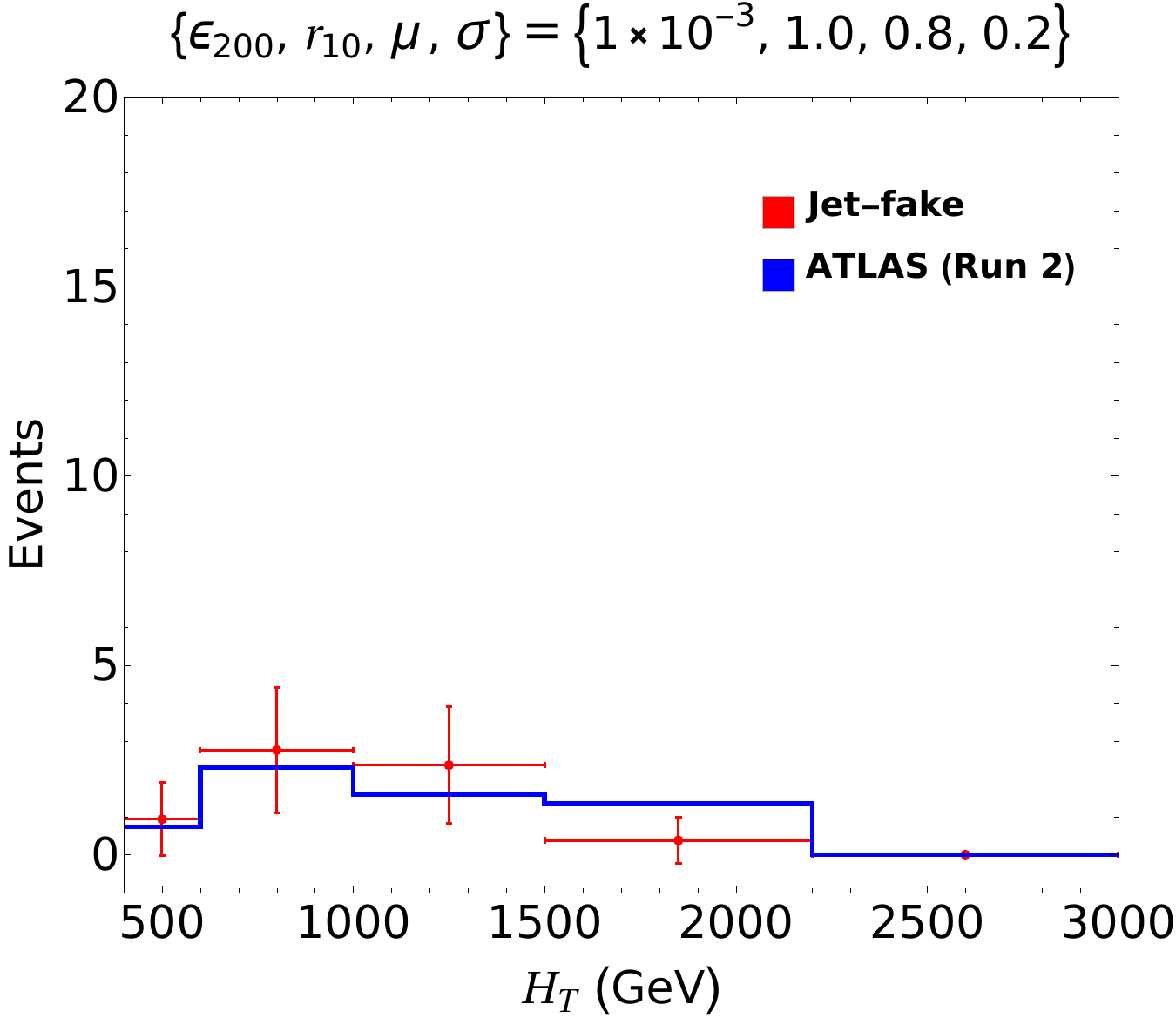}
  \hfill
  \includegraphics[width=0.32\textwidth]{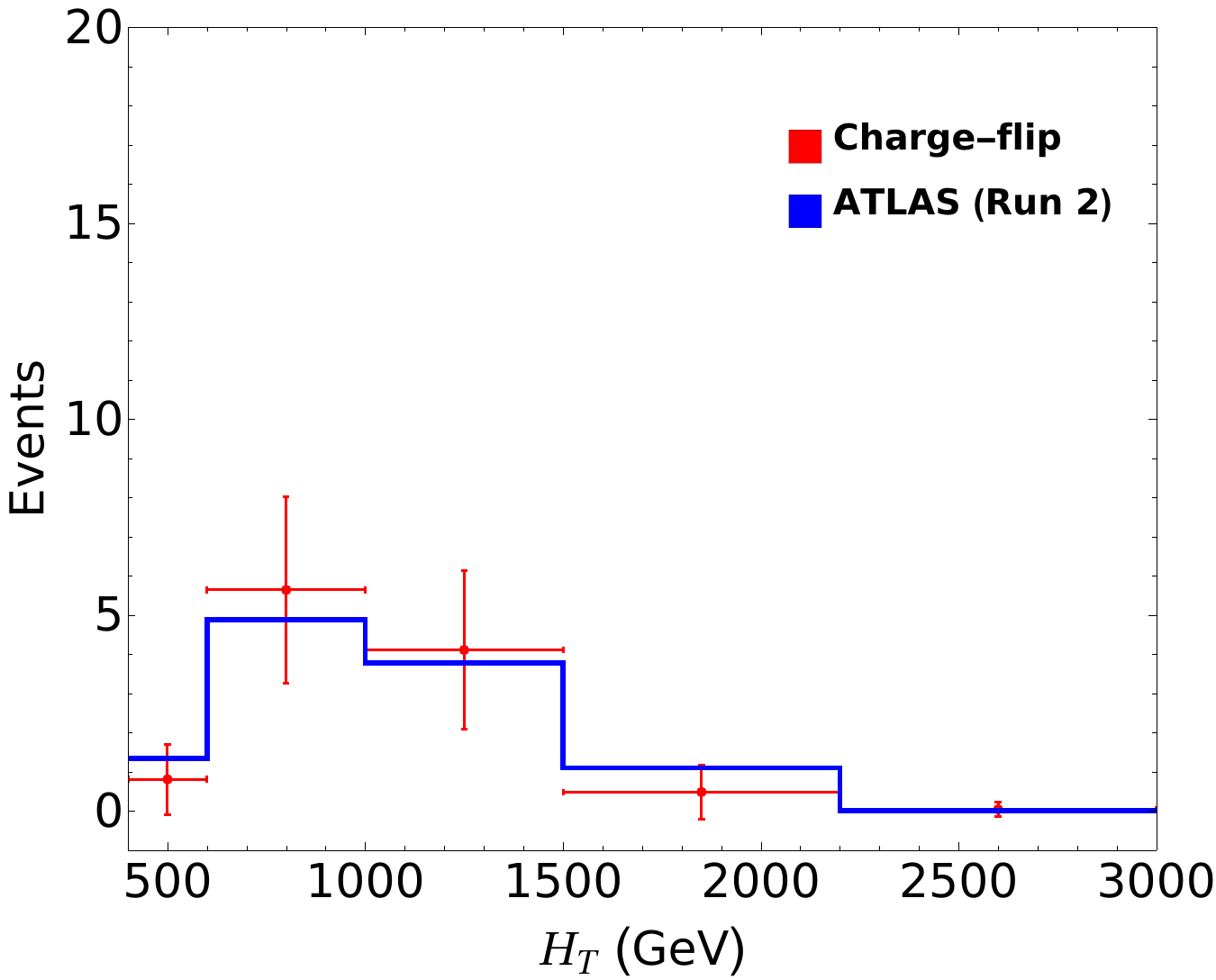}
  \caption{Comparison of the $H_T$ distributions in the $e^\pm e^\pm$ channel between the simulated backgrounds (red) and the ATLAS results (blue) from Fig.~7 of Ref.~\cite{ATLAS:2023cjo}. The left panel shows the multiboson (i.e., prompt) backgrounds; the middle and right panels show the jet‑fake and charge‑flip backgrounds, respectively. The error bars in each bin denote the statistical uncertainties $\sqrt{N_i}$, where $N_i$ is the number of events in the $i$-th bin. The fitted parameters are $\left\{\epsilon_{200}, r_{10}, \mu, \sigma\right\}=\left\{ 10^{-3}, 1.0,0.8,0.2\right\}$.}
  \label{fig:bg_valid}
\end{figure*}

Using Eq.~\eqref{eq:atlas_selection} and including $K$-factors for SM processes~\cite{Chen:2025nzn}, we compare the differential distributions of the SM backgrounds in the $e^\pm e^\pm$ and $\mu^\pm \mu^\pm$ channels with those reported by ATLAS. As shown in Fig.~\ref{fig:bg_valid} for the $e^\pm e^\pm$ channel, the $H_T$ distributions\,\footnote{As additional checks on the background estimation, comparisons with the earlier experimental analyses~\cite{CMS:2017tec,ATLAS:2017xqs,CMS:2018jxx} are performed and also show consistency.} for the three dominant background categories are consistent with the ATLAS results in both shape and normalization, using the fitted parameters $\left\{\epsilon_{200}, r_{10}, \mu, \sigma\right\}=\left\{ 10^{-3}, 1.0,0.8,0.2\right\}$. Similar consistencies are observed for the $\mu^\pm\mu^\pm$ channel, where we obtain the fitted parameters $\left\{\epsilon_{200}, r_{10}, \mu, \sigma\right\}=\left\{ 4\times 10^{-5}, 1.0,0.8,0.2\right\}$. The systematic uncertainty associated with the modeling of fake-lepton backgrounds in the JF method is moderate, as indicated by Ref.~\cite{Curtin:2013zua}, which found an $\mathcal{O}(10\%)$ variation when these parameters were varied.

The validation of our modeling methods in the same-flavor channels $(ee,\mu\mu)$ corroborates the reliability of the simulation across different flavors. Thus, it is straightforward to simulate SM backgrounds in the $e\mu$ channel. 
Furthermore, instead of applying selection cuts sequentially in the cut-based analysis~\cite{ATLAS:2023cjo}, we will explore LHC sensitivities to the masses of $W_R$ and $N_R$ and the right-handed lepton flavor mixing using machine learning.

We emphasize that the SM backgrounds considered in this work are those for the resolved analysis, where two charged leptons are reconstructed. In the ATLAS analyses~\cite{ATLAS:2023cjo}, the boosted channel is also studied and is particularly effective in the regime $m_{W_R} \gg m_{N_R}$. By contrast, for $m_{N_R} \gtrsim 2.5~\text{TeV}$, the resolved analysis yields stronger bounds than the boosted analysis. While lepton flavor effects could in principle be incorporated into the boosted channel, such an analysis would require a dedicated treatment and lies beyond the scope of this work. Hence, both the comparison with the ATLAS limits and our flavor-dependent results, including the HL-LHC projected sensitivities, are restricted to the resolved topology.

% --------------------------------------
\subsection{Machine-learning analysis}
\label{sec:ML-analysis}
% --------------------------------------

In high-energy experiments, conventional 
cut-based analysis
relies on sequential cuts applied to a limited set of kinematic variables, which may overlook complex correlations between observables. To more fully exploit the information contained in collision events, we employ a deep neural network (DNN)~\cite{LeCun:2015pmr,Baldi:2014kfa,deOliveira:2015xxd} as a classifier to distinguish rare signal events from SM background events.

A DNN is a class of artificial neural networks with multiple hidden layers~\cite{LeCun:2015pmr}. In collider analyses, it takes kinematic observables as inputs and learns nonlinear representations that serve as powerful discriminants between signal and background. After training via backpropagation, a selection on the output score yields a signal-enriched sample. Owing to their ability to capture complex correlations in high-dimensional phase space, DNNs outperform conventional cut-based approaches, as is well established in the literature (see Refs.~\cite{Baldi:2014kfa,deOliveira:2015xxd,Guest:2018yhq}).

We consider the following variables as training inputs~\cite{ATLAS:2023cjo,Chen:2025nzn}:
\begin{equation}
p_{T_{j_1}}\,,\ p_{T_{j_2}}\,,\ p_{T_{\ell_1}}\,,\ m_{jj}\,,\  m_{\ell\ell}\,,\ H_T\,,\ \Delta R (\ell_1,\ell_2)\,,
\label{eq:inputs}
\end{equation}
using the same notation as in Eq.~\eqref{eq:atlas_selection}. Their distributions, shown in Fig.~\ref{fig:distribution}, demonstrate that signal events have harder spectra and longer tails than backgrounds — particularly for the heavier mass benchmark — providing a solid basis for discrimination.

Given the wide range in $(m_{W_R}, m_{N_R})$, we divide the parameter space into two regions: low-mass region ($m_{W_R} \leq 3$~TeV) and high-mass region ($m_{W_R} > 3$~TeV). We consider two representative benchmark points:
\begin{itemize}
    \item  BP1: $m_{W_R} = 2$~TeV\,, \quad $m_{N_R} = 0.5$~TeV\,;
    \item  BP2: $m_{W_R} = 4$~TeV\,, \quad $m_{N_R} = 2$~TeV\,.
\end{itemize}
Two separate DNN classifiers are trained for the low- and high-mass regimes, using BP1 and BP2, respectively, with an identical set of input variables.
To incorporate flavor effects, we adopt the benchmark $V_R = V_{\theta=\pi/4}$, where the mixing matrix in the $e\text{--}\mu$ sector is defined as
\begin{equation}
V_{\theta} =
\begin{pmatrix}
\cos\theta & -\sin\theta & 0 \\
\sin\theta & \cos\theta & 0 \\
0 & 0 & 1
\end{pmatrix}\,.
\label{eq:maximal-mixing}
\end{equation}
Nevertheless, the same selection efficiencies apply to other mixing patterns, as the selection criteria are determined by the masses $m_{W_R}$ and $m_{N_R}$. Details of the DNN implementation and validation are provided in Appendix~\ref{app:DNN}.

\begin{figure*}
  \centering
  \includegraphics[width=0.4\textwidth]{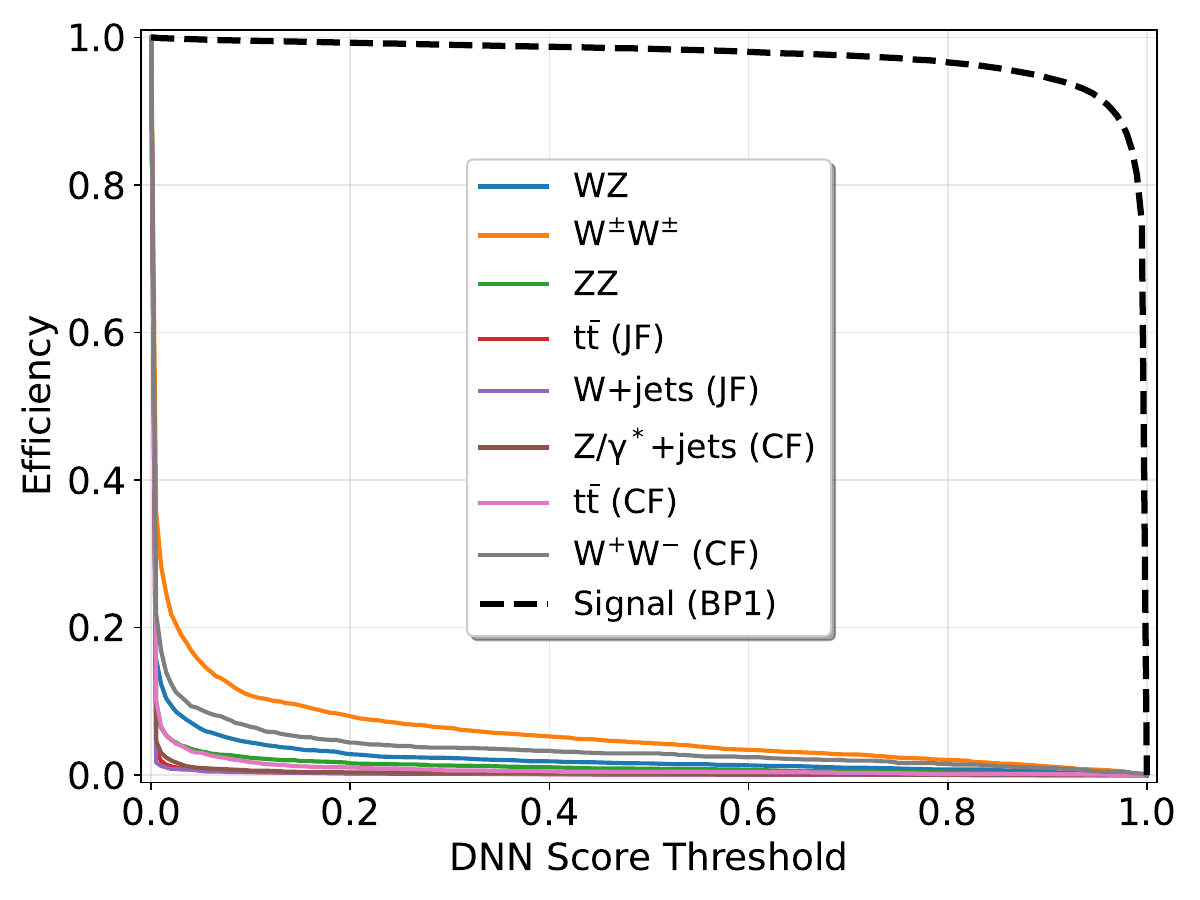}
  \hspace{0.3cm}
  \includegraphics[width=0.4\textwidth]{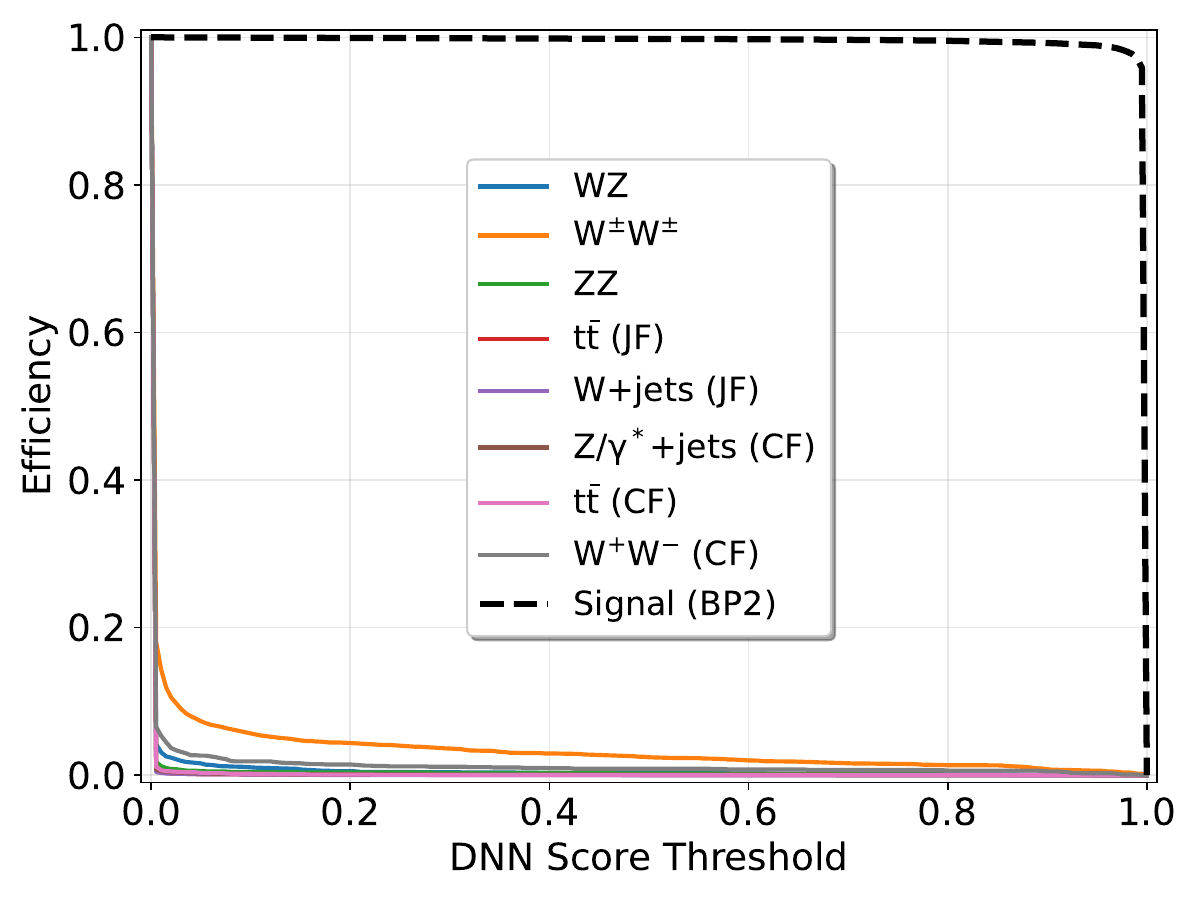}
  \caption{Signal and background efficiencies for the SS dilepton searches as a function of the DNN score threshold. The left and right panels correspond to the low‑mass and high‑mass benchmarks, respectively. 
  }
  \label{fig:DNN-eff_SS}
\end{figure*}

Figure~\ref{fig:DNN-eff_SS} shows the selection efficiencies as a function of the DNN score threshold for the low‑mass and high‑mass models in the SS dilepton searches. The jet‑fake and $Z/\gamma^*$+jets backgrounds are efficiently suppressed at low thresholds due to their softer kinematics, while the $W^\pm W^\pm$, $W^+W^-$ (CF), and $WZ$ backgrounds have harder spectra and therefore larger surviving fractions. The signal efficiency for the high‑mass model decreases slowly up to a threshold that is higher than for the low‑mass model, while background efficiencies drop rapidly, resulting in a clean separation.

To establish the optimal selection criteria and derive exclusion limits, the statistical significance $Z$ is evaluated using the asymptotic profile likelihood ratio test for the signal hypothesis~\cite{Cowan:2010js,Bhattiprolu:2020mwi}
\begin{equation}
\label{eq:significance}
Z = \sqrt{ 2 \left( \mathcal{S} - \mathcal{B} \ln\left(\frac{\mathcal{S} + \mathcal{B}}{\mathcal{B}}\right) \right)}\,,
\end{equation}
where $\mathcal{S}$ and $\mathcal{B}$ denote the number of signal and background events remaining after the DNN selection, respectively.
The number of SS signal events in each channel after the optimal DNN selection is given in Eq.~\eqref{eq:events_SS}. For the $\ell^\pm \ell^\pm$ channel, the significance is computed by combining the events from the $e^\pm e^\pm$, $\mu^\pm \mu^\pm$, and $e^\pm \mu^\pm$ channels, with $\mathcal{S} = N_{e^\pm e^\pm} + N_{\mu^\pm \mu^\pm} + N_{e^\pm \mu^\pm}$, while $\mathcal{B}$ includes all SM background contributions with a pair of SS charged leptons~\cite{CMS:2018jxx}.

\begin{table}[!htbp]
\small
\setlength{\tabcolsep}{3pt}
\caption{Cross sections for the SS dilepton signal and background processes in the case of low-mass model. Values are shown at three stages: after the basic cuts ($\sigma_{\text{basic}}$), after the baseline selection ($\sigma_{\ell\ell jj}$), and after the optimal DNN score selection ($\sigma_{\text{DNN}}$). The bottom rows provide the comparison of the total background and signal cross sections. }
\label{tab:cutcross_low}
\begin{ruledtabular}
\begin{tabular}{lccc}
Process & $\sigma_{\text{basic}}$ (pb) & $\sigma_{\ell\ell jj}$ (pb) & $\sigma_{\text{DNN}}$ (pb) \\
\hline
$WZ$ & $4.31\!\times\!10^{-1}$ & $1.93\!\times\!10^{-2}$ & $7.74\!\times\!10^{-5}$ \\
$W^\pm W^\pm$ & $1.17\!\times\!10^{-2}$ & $2.59\!\times\!10^{-3}$ & $3.54\!\times\!10^{-5}$ \\
$ZZ$ & $4.15\!\times\!10^{-2}$ & $1.75\!\times\!10^{-3}$ & $3.15\!\times\!10^{-6}$ \\
$t\bar{t}$ (JF) & $1.08\!\times\!10^{-1}$ & $5.33\!\times\!10^{-3}$ & $8.03\!\times\!10^{-6}$ \\
$W+\text{jets}$ (JF) & $9.47\!\times\!10^{-1}$ & $2.14\!\times\!10^{-2}$ & $2.96\!\times\!10^{-5}$ \\
$t\bar{t}$ (CF) & $1.17\!\times\!10^{-1}$ & $2.42\!\times\!10^{-2}$ & $5.47\!\times\!10^{-6}$ \\
$Z/\gamma^*+\text{jets}$ (CF) & $2.71\!\times\!10^{-1}$ & $1.27\!\times\!10^{-1}$ & $1.39\!\times\!10^{-5}$ \\
$W^+W^-$ (CF) & $1.11\!\times\!10^{-3}$ & $2.28\!\times\!10^{-4}$ & $2.53\!\times\!10^{-6}$ \\
\hline
Total bkg & $1.93$ & $2.02\!\times\!10^{-1}$ & $1.75\!\times\!10^{-4}$ \\
Signal & $5.29\!\times\!10^{-2}$ & $2.49\!\times\!10^{-2}$ & $2.36\!\times\!10^{-2}$ \\
\end{tabular}
\end{ruledtabular}
\end{table}

By maximizing the expected significance $Z$, we obtain optimal DNN score thresholds of 0.91 for the low-mass model and 0.99 for the high-mass model in the SS dilepton searches. Table~\ref{tab:cutcross_low} summarizes the cross sections for the signal and SM backgrounds at different selection stages for the low-mass model. The three stages are: $\sigma_{\text{basic}}$ after applying the basic cuts at the generator level (cf. Eq.~\eqref{eq:basic_cuts}), $\sigma_{\ell\ell jj}$ after imposing the baseline selection, and $\sigma_{\text{DNN}}$ after requiring the optimal threshold of 0.91. The signal cross section remains robust throughout the selections, while the total background is suppressed by nearly four orders of magnitude relative to the generator level, leading to a significantly enhanced signal-to-background ratio.

\begin{table}[!htbp]
\small
\setlength{\tabcolsep}{3pt}
\caption{Event yields after the optimal DNN score selection for the low-mass model. The backgrounds (individual categories and total) and the SS dilepton signal for BP1 are shown for the three SS dilepton channels. A dash (--) indicates that the corresponding background contribution is negligible. The last row gives the signal efficiency in each channel.
}
\label{tab:cutevent_low}
\centering
\begin{tabular}{lccc}
\hline
Process & $e^\pm \mu^\pm$ & $e^\pm e^\pm$ & $\mu^\pm \mu^\pm$ \\
\hline
Multiboson & 6 & 4 & 6 \\
Jet-fake & 3 & 3 & -- \\
Charge-flip & -- & 3 & -- \\
\hline
Total bkg & 9 & 10 & 6 \\
Signal & 1615 & 641 & 1027 \\
Eff [\%] & 44 & 35 & 56 \\
\hline
\end{tabular}
\end{table}

To evaluate flavor-specific sensitivities, Table~\ref{tab:cutevent_low} provides the event yields after the optimal DNN selection, which correspond to an integrated luminosity of $139~\text{fb}^{-1}$ at LHC Run~2. The last row lists the corresponding selection efficiencies, i.e., the fraction of events that survive the DNN selection: $\epsilon_{\mu\mu}=56\%$, $\epsilon_{e\mu}=44\%$, and $\epsilon_{ee}=35\%$. Thus, the $\mu^\pm\mu^\pm$ channel has the highest efficiency, followed by $e^\pm\mu^\pm$ and $e^\pm e^\pm$.
This difference in efficiencies is primarily attributed to the better reconstruction performance of muons compared to electrons at the LHC. 
We note that the jet‑fake backgrounds in the $e^\pm \mu^\pm$ channel arise primarily from $j\to e$, rather than from $j\to \mu$.

\begin{table}[H]
\small
\setlength{\tabcolsep}{3pt}
\caption{Same as Table~\ref{tab:cutcross_low}, but for the high-mass model with the optimal DNN score threshold of 0.99. After the DNN selection, only the $WZ$ and $W^\pm W^\pm$ backgrounds are non‑negligible; all other backgrounds are suppressed to $\lesssim 10^{-7}$~pb or below.}
\label{tab:cutcross_high}
\centering
\begin{tabular}{lccc}
\hline
Process & $\sigma_{\text{basic}}$ (pb) & $\sigma_{\ell\ell jj}$ (pb) & $\sigma_{\text{DNN}}$ (pb) \\
\hline
$WZ$ & $4.31 \!\times\! 10^{-1}$ & $1.93 \!\times\! 10^{-2}$ & $4.18 \!\times\! 10^{-6}$ \\
$W^\pm W^\pm $ & $1.17 \!\times\! 10^{-2}$ & $2.59 \!\times\! 10^{-3}$ & $6.64 \!\times\! 10^{-6}$ \\
$ZZ$ & $4.15\!\times\!10^{-2}$ & $1.75\!\times\!10^{-3}$ & $\lesssim 10^{-7}$ \\
$t\bar{t}$ (JF) & $1.08\!\times\!10^{-1}$ & $5.33\!\times\!10^{-3}$ & $\lesssim 10^{-7}$ \\
$W+\text{jets}$ (JF) & $9.47\!\times\!10^{-1}$ & $2.14\!\times\!10^{-2}$ & $\lesssim 10^{-7}$ \\
$t\bar{t}$ (CF) & $1.17\!\times\!10^{-1}$ & $2.42\!\times\!10^{-2}$ & $\lesssim 10^{-8}$ \\
$Z/\gamma^*+\text{jets}$ (CF) & $2.71\!\times\!10^{-1}$ & $1.27\!\times\!10^{-1}$ & $\lesssim 10^{-7}$ \\
$W^+W^-$ (CF) & $1.11\!\times\!10^{-3}$ & $2.28\!\times\!10^{-4}$ & $\lesssim 10^{-8}$ \\
\hline
Total bkg & $1.93$ & $2.02\!\times\!10^{-1}$ & $1.08 \!\times\! 10^{-5}$ \\
Signal & $4.24 \!\times\! 10^{-4}$ & $2.30 \!\times\! 10^{-4}$ & $2.23 \!\times\! 10^{-4}$ \\
\hline
\end{tabular}
\end{table}

\begin{table}[H]
\small
\setlength{\tabcolsep}{3pt}
\caption{Same as Table~\ref{tab:cutevent_low}, but for the high-mass model with the optimal DNN score threshold of 0.99. 
}
\label{tab:cutevent_high}
\centering
\begin{tabular}{lccc}
\hline
Process & $e^\pm \mu^\pm$ & $e^\pm e^\pm$ & $\mu^\pm \mu^\pm$ \\
\hline
Multiboson & 0.67 & 0.49 & 0.34 \\
\hline
Total bkg & 0.67 & 0.49 & 0.34 \\
Signal & 15.37 & 6.44 & 9.17 \\
Eff [\%] & 52.13 & 43.58 & 62.16 \\
\hline
\end{tabular}
\end{table}

Table~\ref{tab:cutcross_high} shows the cross sections at each stage for the high‑mass model in the SS dilepton channel. After the DNN selection, only the $WZ$ and $W^\pm W^\pm$ backgrounds have non‑negligible cross sections; all other backgrounds are suppressed to $\lesssim 10^{-7}$~pb or smaller. Table~\ref{tab:cutevent_high} presents the corresponding event yields in the three SS dilepton channels. Compared to the low‑mass case (cf. Table~\ref{tab:cutevent_low}), the high‑mass model yields significantly fewer background events after the DNN selection\,\footnote{For an integrated luminosity of $139~\text{fb}^{-1}$, some expected event yields are below unity; we show all yields with two decimal places for uniform precision.}.

An analogous procedure is applied to the OS dilepton channels, as detailed in Appendix~\ref{app:ML_OS}. Although the OS channels are subject to significantly larger backgrounds at the generator level, the DNN selection effectively suppresses these contributions, making them comparable to those in the SS dilepton channels. Consequently, the resulting signal significance in the $\ell^+ \ell^-$ channel for the low-mass (high-mass) benchmarks is $Z = 77.3\ (6.3)$, which is close to $Z = 79.5\ (7.3)$ achieved in the SS dilepton channels.

We perform a parameter scan over the $m_{W_R}\text{--}m_{N_R}$ plane for a given right-handed lepton mixing matrix $V_R$, dividing the analysis into low- and high-mass regimes. For the low-mass regime we use the low-mass model; for the high-mass regime we use a model optimized to exploit the distinct kinematics of heavy $W_R$ and $N_R$ decays. For each parameter point, we compute the SS and OS significances using Eq.~\eqref{eq:significance}. The combined significance for the SS and OS channels (denoted as SS+OS) is then obtained by taking the square root of the sum of the squares of the individual significances~\cite{Cowan:2010js}. Exclusion limits at 95\% C.L. for the SS, OS, and SS+OS analyses are derived by requiring the corresponding significance to reach 1.96.

%%%%%%%%%%%%%%%%%%%%%%%%%%%%%%%%%%%%%%%%%
\section{LHC sensitivities}
\label{sec:LHC_sensitivities}
%%%%%%%%%%%%%%%%%%%%%%%%%%%%%%%%%%%%%%%%%

In this section, we evaluate the sensitivity to the Keung-Senjanović process at LHC Run 2, as well as the projected reach at the HL-LHC. Both the sensitivities from SS dilepton searches and their combination with OS dilepton searches are investigated. 

To assess the impact of flavor structure of the right-handed lepton mixing matrix, we consider three benchmark mixing scenarios:
\begin{itemize}
\item \textbf{Unmixed}: $V_{e1}=1$ or $V_{\mu1}=1$, representing the case of no lepton flavor mixing.
\item \textbf{Maximal mixing}: $V_R = V_{\theta = \pi/4}$, corresponding to maximal mixing in the $e\text{--}\mu$ sector.
\item \textbf{PMNS-like mixing}: $V_R = V_{\rm PMNS}^{(*)}$, as commonly adopted in the minimal LRSM with generalized parity and charge conjugation~\cite{deVries:2022nyh}.
\end{itemize}
The signal event yields for the $ee$ and $\mu\mu$ channels scale as $|V_{e1}|^4$ and $|V_{\mu1}|^4$, respectively. For maximal mixing, $|V_{e1}| = |V_{\mu1}| = 0.707$. For the PMNS‑like pattern, the latest global fits for normal mass ordering give $|V_{e1}| = 0.82$ and $|V_{\mu1}| = 0.33$ (central values)~\cite{Esteban:2024eli,JUNO:2025gmd,Esteban:2026phq}. Consequently, while the scaling factors are symmetric under maximal mixing, the PMNS‑like pattern leads to a significantly enhanced $ee$ yield at the expense of a suppressed $\mu\mu$ contribution.

% --------------------------------------
\subsection{Exclusion limits on \tf{$W_R$}{WR} and \tf{$N_R$}{NR} masses}
\label{sec:exclusion_mass}
% --------------------------------------

\begin{figure}[!htbp]
  \centering
  \includegraphics[width=0.23\textwidth]{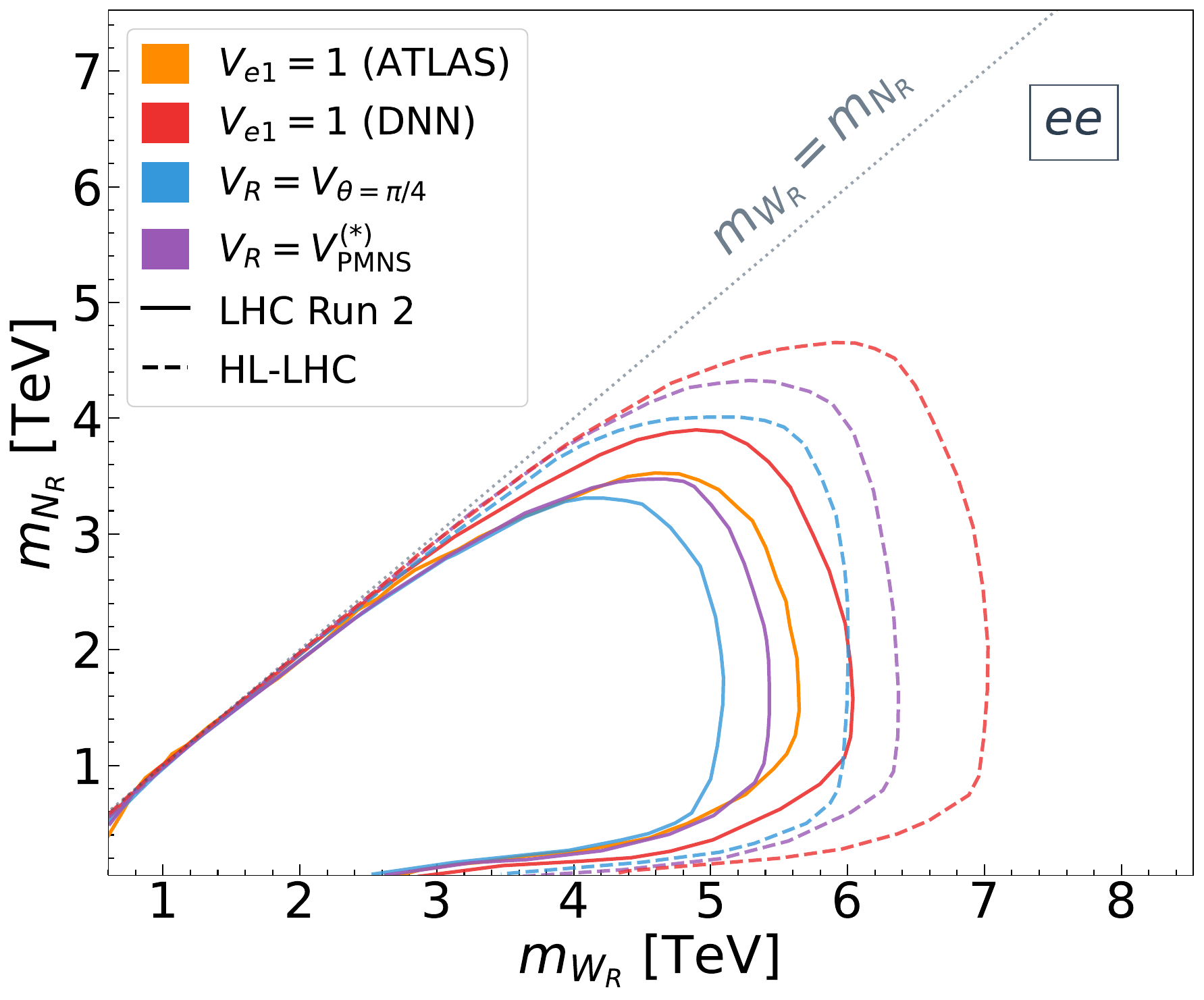}
  \hspace{0.1cm}
  \includegraphics[width=0.23\textwidth]{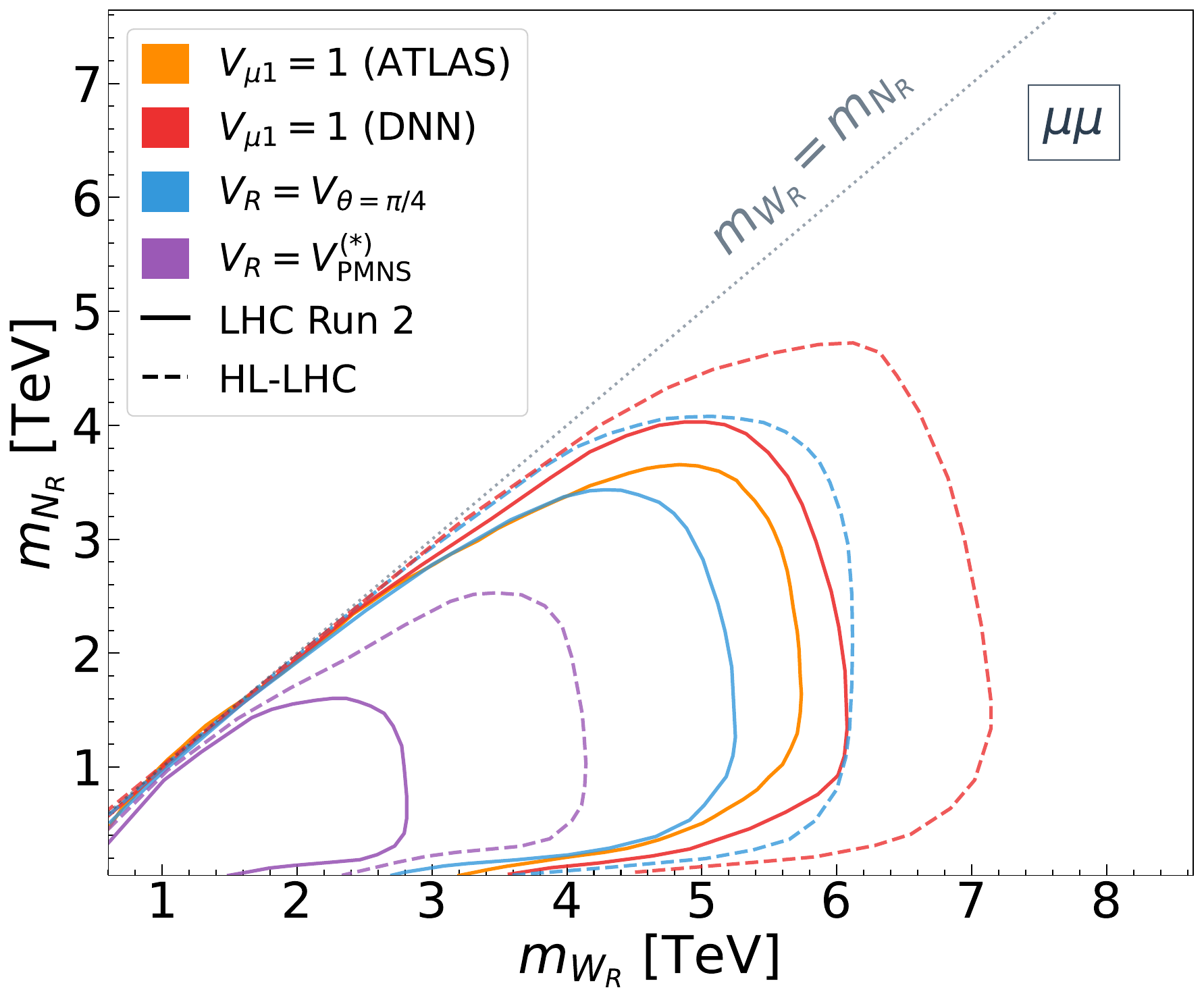} \\ [0.1cm]
  \includegraphics[width=0.23\textwidth]{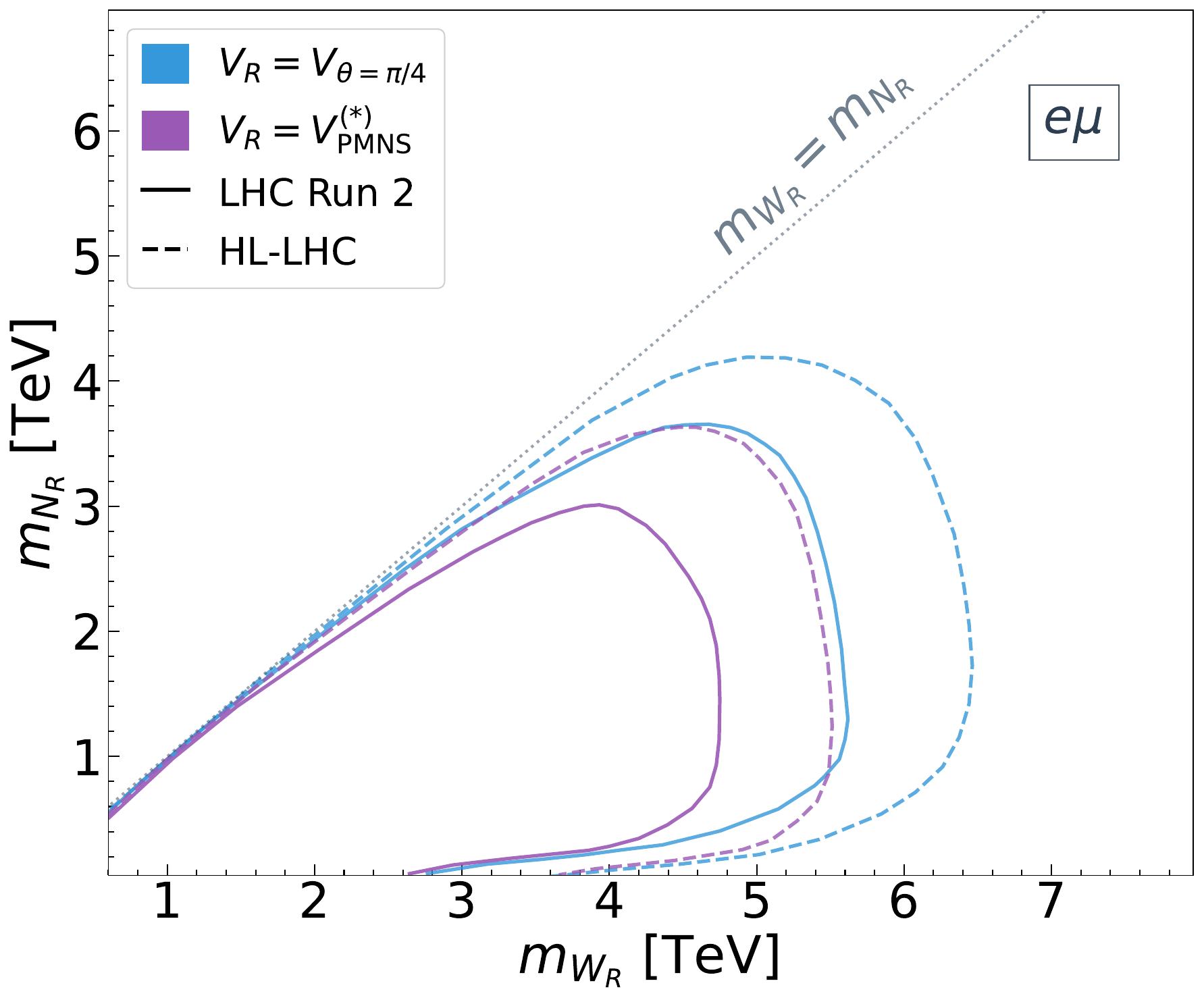}
  \hspace{0.1cm}
  \includegraphics[width=0.23\textwidth]{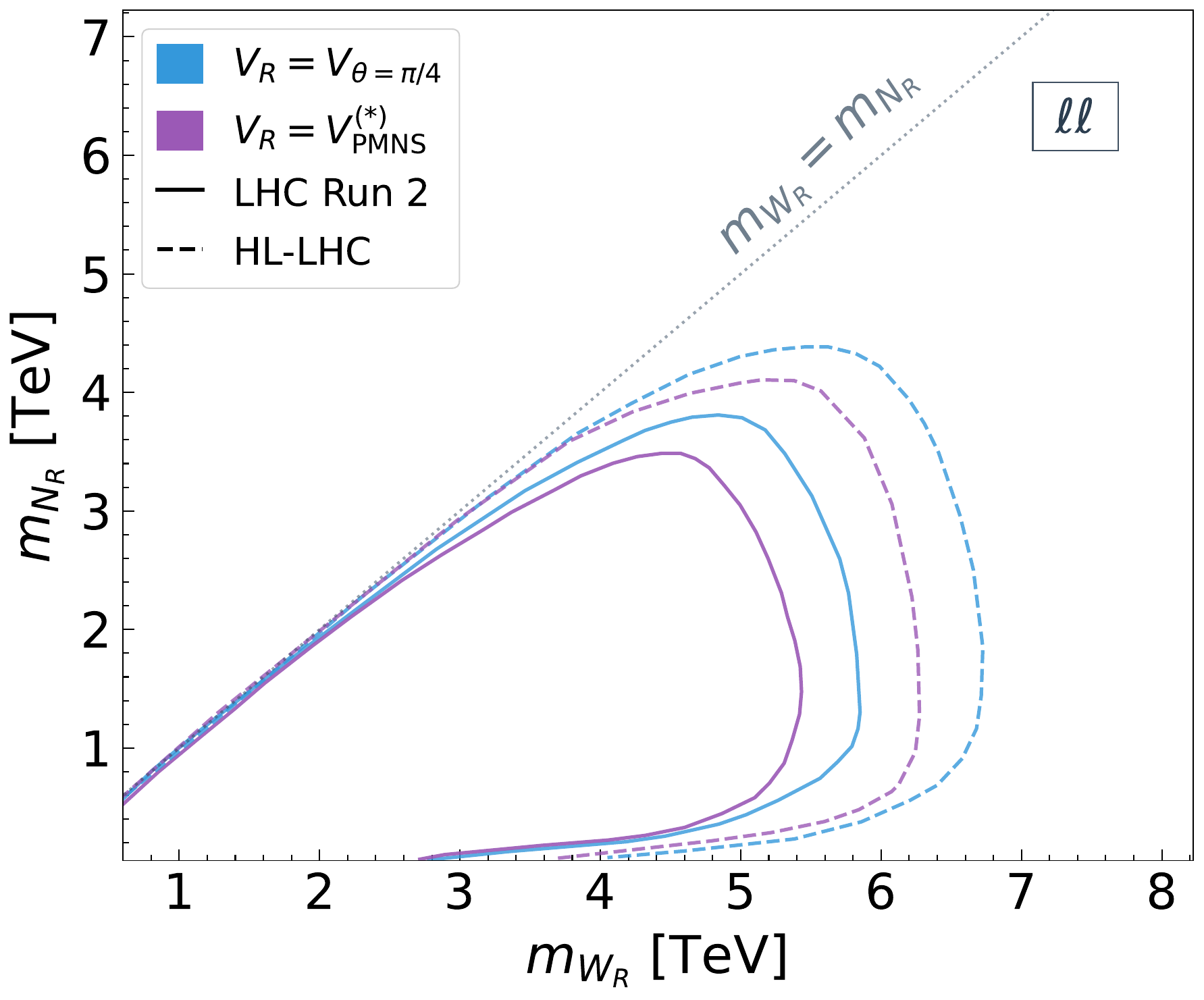}
  \caption{95\% C.L. constraints on $m_{W_R}$ and $m_{N_R}$ from combined SS and OS dilepton searches at LHC Run 2 (solid) and HL-LHC (dashed). The panels correspond to the $ee$ (upper left), $\mu\mu$ (upper right), $e\mu$ (lower left), and combined $\ell\ell$ ($\ell=e,\mu$) (lower right) channels. The blue and purple contours represent the maximal-mixing scenario ($V_R=V_{\theta=\pi/4}$) and the PMNS-like scenario ($V_R=V_{\mathrm{PMNS}}^{(*)}$), respectively. In the $ee$ and $\mu\mu$ channels, the red contours indicate the unmixed cases ($V_{e1}=1$ or $V_{\mu1}=1$) obtained with the DNN analysis, while the orange contours indicate the ATLAS results. The gray dotted line in each panel marks the kinematic threshold $m_{W_R}=m_{N_R}$. 
  }
  \label{fig:mass_reach_SS+OS}
\end{figure}

The sensitivity of individual channels in both SS and OS dilepton searches depends strongly on the mixing benchmark. In the $ee$ and $\mu\mu$ channels, the signal cross sections scale with $|V_{e1}|^4$ and $|V_{\mu1}|^4$, respectively, yielding the strongest limits in the unmixed scenario. For the maximal-mixing scenario, the $e\mu$ channel has approximately twice the event yield of the $ee$ or $\mu\mu$ channels (cf. Eqs.~\eqref{eq:events_SS} and~\eqref{eq:events_OS}), making it the most sensitive individual channel. In contrast, for the PMNS-like scenario, the sensitivities follow the order $ee > e\mu > \mu\mu$, given the mixing entries.

The combined 95\% C.L. constraints from the SS+OS dilepton searches are shown in Fig.~\ref{fig:mass_reach_SS+OS} (see Fig.~\ref{fig:mass_reach_SS} for those from SS dilepton searches). 
The solid and dashed curves show the LHC Run~2 results and the HL-LHC projections, respectively.
The orange curves correspond to the exclusion limits reported by ATLAS in the resolved analysis of the Run 2 data~\cite{ATLAS:2023cjo}\,\footnote{The ATLAS analysis derives upper limits by combining all signal regions without distinguishing between the SS and OS channels. Our limits are obtained using two‑sided confidence intervals and without a binned likelihood analysis, both of which lead to a conservative estimate of the sensitivity of the DNN analysis.}. For the unmixed $ee$ and $\mu\mu$ channels, 
the expected sensitivities
from our DNN analysis (red curves) are stronger than the corresponding ATLAS results, owing to improved signal efficiency and background rejection in the ML approach.

We further combine all flavor contributions in the $\ell\ell$ channel ($\ell = e,\mu$), where the total signal cross section scales as $(|V_{e1}|^2 + |V_{\mu1}|^2)^2$. Thus, flavor mixing mainly redistributes the signal among the different dilepton final states. In the maximal-mixing scenario ($|V_{e1}| = |V_{\mu1}| = 0.707$), the total signal cross section coincides with that of either unmixed case ($V_{e1}=1$ or $V_{\mu1}=1$) before any selection cuts. However, the combined $\ell\ell$ channel receives background contributions from all three flavor channels, leading to a larger total background yield than in any single-flavor channel. Consequently, the resulting exclusion is weaker than in the individual unmixed cases. In the PMNS-like scenario, $|V_{e1}|^2 + |V_{\mu1}|^2 < 1$, thus the signal rate is further suppressed, yielding limits that are less stringent than those in the maximal-mixing case.

Using the LHC Run 2 results as a reference, we project the sensitivity at the HL-LHC. The increased luminosity substantially extends the accessible mass range. Under maximal (PMNS-like) mixing, the HL-LHC 
is expected to probe
$m_{W_R}$ up to $6.7$ ($6.3$)~TeV for $m_{N_R}=2$~TeV, and $m_{N_R}$ up to $4.4$ ($4.1$)~TeV for $m_{W_R}=5.6$ ($5.4$)~TeV in the $\ell\ell$ channel. These reaches in $m_{W_R}$ and $m_{N_R}$ are parameter-dependent projections of the sensitivity contours in the $(m_{W_R},m_{N_R})$ plane, rather than independent endpoints. Similar improvements are observed in the individual $ee$, $\mu\mu$, and $e\mu$ channels, as shown in Fig.~\ref{fig:mass_reach_SS+OS}.

% --------------------------------------
\subsection{Bounds on right-handed lepton flavor mixing}
% --------------------------------------

We now investigate bounds on the mixing entries $|V_{e1}|$ and $|V_{\mu1}|$ at LHC Run 2 and the HL-LHC.
The relations in Eqs.~\eqref{eq:events_SS} and~\eqref{eq:events_OS} imply that the LHC bounds on $|V_{e1}|$ and $|V_{\mu1}|$ are independent of the assumption of $V_{\tau1}$, and the allowed region in the $|V_{e1}|\text{--}|V_{\mu1}|$ plane lies within a quarter-circle. Different reconstruction efficiencies of electrons and muons slightly distort this boundary, giving a mild asymmetry.

\begin{figure}[!htbp]
  \centering
  \includegraphics[width=0.25\textwidth]{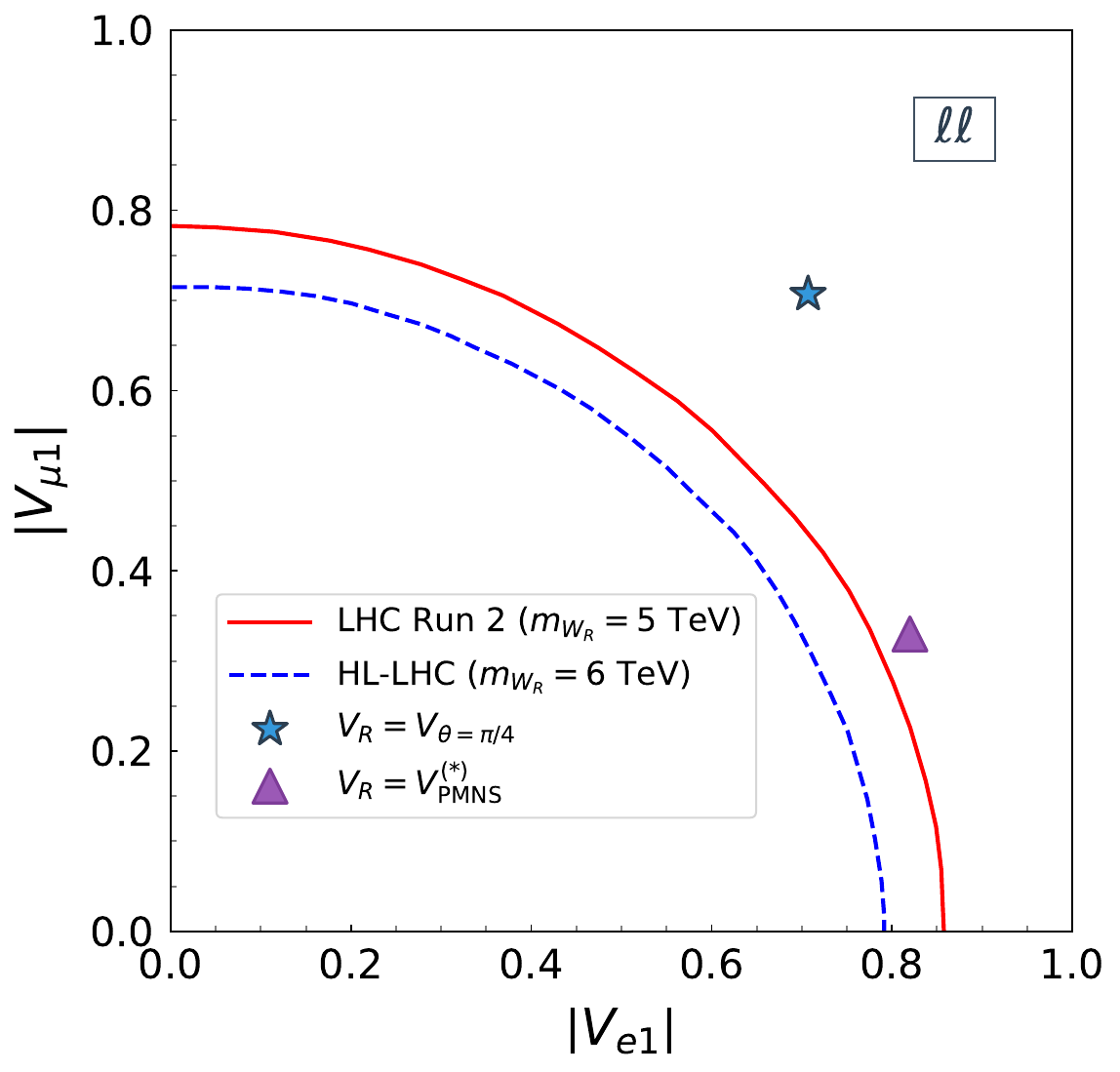}
  \caption{95\% C.L. constraints on $|V_{e1}|$ and $|V_{\mu1}|$ for fixed $m_{N_R}=3$~TeV in the combined SS+OS dilepton channels. The red solid contour shows the limit from LHC Run 2 for $m_{W_R}=5$~TeV, while the blue dashed contour indicates the HL-LHC sensitivity for $m_{W_R}=6$~TeV. The star and triangle mark the values of $|V_{e1}|$ and $|V_{\mu1}|$ for the maximal-mixing and PMNS-like scenarios, respectively.}
  \label{fig:mixing_bound_SS+OS}
\end{figure}

Figure~\ref{fig:mixing_bound_SS+OS} shows the constraints in the $|V_{e1}|\text{--}|V_{\mu1}|$ plane for two benchmark points: $(m_{W_R}, m_{N_R}) = (5~\text{TeV}, 3~\text{TeV})$ at LHC Run~2 and $(6~\text{TeV}, 3~\text{TeV})$ at the HL‑LHC.
For the LHC Run~2 benchmark, the 
sensitivity contour
is characterized by $\sqrt{|V_{e1}|^2 + |V_{\mu1}|^2} \sim 0.8\text{--}0.85$.
For the HL‑LHC benchmark, the increased center‑of‑mass energy and integrated luminosity compensate for the impact of the larger $m_{W_R}$. Consequently, the sensitivity extends to smaller values of $|V_{e1}|$ and $|V_{\mu1}|$, reaching $\sqrt{|V_{e1}|^2 + |V_{\mu1}|^2} \sim 0.7\text{--}0.8$, thereby probing a larger portion of the flavor mixing plane.
In particular, by combining the SS and OS searches, our simulated analysis has the expected sensitivity to exclude the maximal‑mixing and PMNS‑like benchmark points for the chosen masses.

%%%%%%%%%%%%%%%%%%%%%%%%%%%%%%%%%%%%%%%%%
\section{Complementarities with CLFV searches}
\label{sec:LHC_CLFV}
%%%%%%%%%%%%%%%%%%%%%%%%%%%%%%%%%%%%%%%%%

The CLFV processes $\mu \rightarrow e\gamma$, $\mu \rightarrow eee$ and $\mu \rightarrow e$ conversion in nuclei are highly suppressed in the SM. In the LRSM, these processes receive additional contributions from right-handed charged currents as well as from the Yukawa interactions between charged leptons and the right-handed triplet scalars (cf. Eq.~\eqref{eq:yukawa}). These contributions have been extensively studied in Refs.~\cite{Mohapatra:1980yp,Cirigliano:2004mv,Das:2012ii,Barry:2013xxa,Deppisch:2014zta,Bambhaniya:2015ipg,Borah:2016iqd}. Representative Feynman diagrams are shown in Appendix~\ref{app:LFV}.

\begin{figure}[!htbp]
  \centering
  \includegraphics[width=0.35\textwidth]{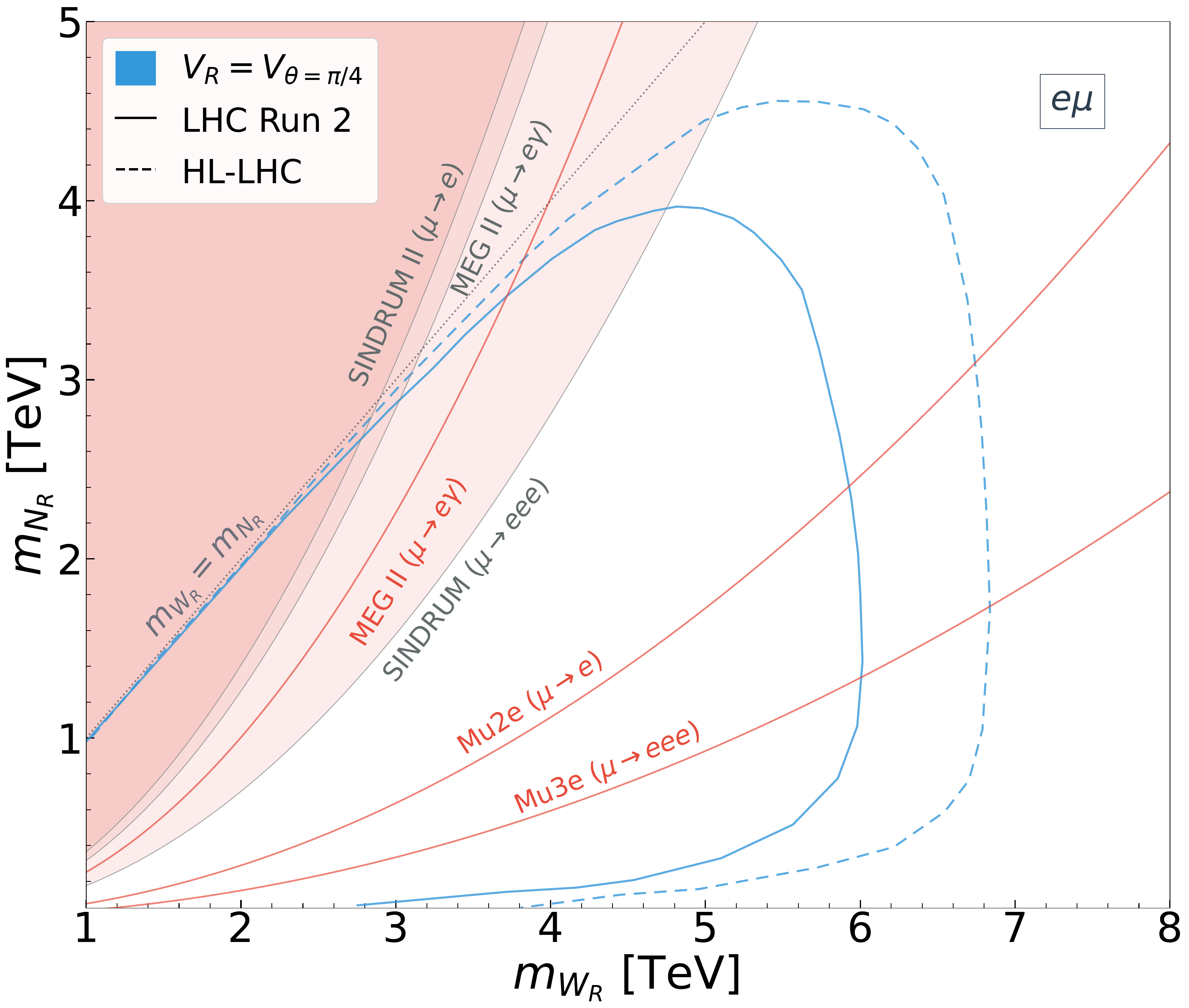}
  \caption{Comparison of LHC and CLFV constraints on $m_{W_R}$ and $m_{N_R} \equiv m_{N_1}$ in the maximal‑mixing scenario, assuming $m_{\Delta_{L,R}^{--}} = m_{\Delta_L^-} = m_{W_R}$ and $(m_{N_2}-m_{N_1})/m_{N_1}=0.01$. Blue solid and dashed curves show 95\% C.L. exclusion limits for the $e\mu+jj$ channel at the LHC and HL‑LHC, respectively. Current 90\% C.L. bounds from CLFV experiments are shown as red shaded regions with gray boundaries, while red curves indicate future projections. The gray dotted line corresponds to $m_{W_R}=m_{N_R}$.}
  \label{fig:LFV_sensitivity}
\end{figure}

The current 90\% C.L. upper limits on the branching ratios are given by
\begin{align}
{\rm Br}(\mu \rightarrow e \gamma)
&< 1.5 \!\times\! 10^{-13}
\quad \text{(MEG II~\cite{MEGII:2025gzr})}\,,\\
{\rm Br}(\mu \rightarrow eee)
&< 1.0 \!\times\! 10^{-12}
\quad \text{(SINDRUM~\cite{SINDRUM:1987nra})}\,,\\
R_{\rm Au}(\mu \rightarrow e)
&< 7.0 \!\times\! 10^{-13}
\quad \text{(SINDRUM II~\cite{SINDRUMII:2006dvw})}\,.
\end{align}
Future experiments are expected to significantly improve these sensitivities. The projected reaches are
\begin{align}
{\rm Br}(\mu \rightarrow e \gamma)
&\lesssim 6.0 \!\times\! 10^{-14}
\quad \text{(MEG II~\cite{MEGII:2025gzr})}\,,\\
{\rm Br}(\mu \rightarrow eee)
&\lesssim 2.0 \!\times\! 10^{-15}
\quad \text{(Mu3e Phase I~\cite{Mu3e:2020gyw})}\,,\\
R_{\rm Al}(\mu \rightarrow e)
&\lesssim 6.2 \!\times\! 10^{-16}
\quad \text{(Mu2e Run I~\cite{Mu2e:2022ggl})}\,.
\end{align}

In addition to $W_R$ and $N_i$, the charged scalars $\Delta_{L,R}^{\pm\pm}$ and $\Delta_L^{\pm}$ also contribute to the CLFV processes. Assuming the charged scalar masses are degenerate with $W_R$~\cite{Cirigliano:2004mv}, the branching ratios take the form given in Eqs.~\eqref{eq:br_muea}~\eqref{eq:br_mu3e}~\eqref{eq:br_mu2e}.

If only one heavy neutrino contributes, the resulting constraints are stringent. In contrast, when two or more heavy neutrinos are nearly degenerate, large cancellation can occur, significantly suppressing the CLFV branching ratios compared to the individual contributions~\cite{Das:2012ii}. Accordingly, we assume maximal mixing and a relative mass splitting $(m_{N_2}-m_{N_1})/m_{N_1}=0.01$ to facilitate a direct comparison between the mass reaches of collider searches and CLFV experiments.
Under these assumptions, interference effects at the LHC can be neglected~\cite{Das:2012ii,Gluza:2015goa,Das:2017hmg}, and the signal cross section for the Keung–Senjanović process is approximately twice that for a single $N_1$. Such an enhancement is included in the blue LHC sensitivity curves in Fig.~\ref{fig:LFV_sensitivity}. This near-degenerate heavy-neutrino spectrum differs from that assumed in Secs.~\ref{sec:LHC_sensitivities}--\ref{sec:KS-process}, where $N_1$ is well separated from $N_{2,3}$.

Using the current limits and projections of CLFV experiments, we derive constraints on $m_{W_R}$ and $m_{N_R} \equiv m_{N_1}$ for the benchmark $m_{\Delta_{L,R}^{--}} = m_{\Delta_L^-} = m_{W_R}$, as illustrated in Fig.~\ref{fig:LFV_sensitivity}. The existing CLFV bounds (red shaded regions) predominantly constrain the domain $m_{N_R} > m_{W_R}$\,\footnote{The vacuum stability and perturbativity bounds on the right-handed Majorana neutrino masses in the LRSM were investigated in Refs.~\cite{Mohapatra:1986pj,Maiezza:2016bzp}.}. In combination with the LHC Run~2 exclusion limit obtained from the SS+OS searches in the $e\mu+jj$ channel (solid blue curve), these constraints exclude most of the parameter space with $m_{W_R} \lesssim 6~\text{TeV}$, irrespective of the value of $m_{N_R}$. Projected sensitivities of future CLFV experiments (red curves) are expected to cover a large fraction of the region $m_{W_R} > m_{N_R}$. Notably, they may surpass the HL‑LHC reach (dashed blue curve), thereby probing $m_{W_R} \gtrsim 7~\text{TeV}$.

Finally, we note that neutrinoless double beta ($\onbb$) decay provides
another probe of the LRSM. The short-range contribution to the
$\onbb$ decay amplitude scales as
$\sim m_p (m_W/m_{W_R})^4 \sum_i (V_{ei})^2/m_{N_i}$~\cite{Barry:2013xxa}, where $m_p$ denotes the nucleon mass. Due to the inverse dependence on $m_{N_i}$, this contribution is dominated by the lightest of the three right-handed neutrinos. As shown in
Refs.~\cite{Tello:2010am,Li:2020flq}, for TeV-scale $W_R$ and $N_R$ masses, the $\onbb$-decay constraints from current and future tonne-scale experiments are weaker than those from direct LHC searches for the Keung--Senjanović process~\cite{Nemevsek:2018bbt}, and are therefore not included in our combined analysis.

%%%%%%%%%%%%%%%%%%%%%%%%%%%%%%%%%%%%%%%%%
\section{Conclusion}
\label{sec:conclusion}
%%%%%%%%%%%%%%%%%%%%%%%%%%%%%%%%%%%%%%%%%

In this work, we have performed a comprehensive study of the Keung–Senjanović process $pp \to W_R \to \ell_\alpha N_R \to \ell_\alpha \ell_\beta jj$ with $\ell_{\alpha,\beta}=e,\mu$ in the minimal Left–Right Symmetric Model (LRSM) at LHC Run~2 and the HL-LHC, with particular emphasis on the role of right-handed lepton flavor mixing. To enhance signal sensitivity, we employed a deep neural network optimized for different mass regimes, achieving significantly improved signal efficiency and background rejection over conventional cut-based analyses.

The expected sensitivity depends strongly on the mixing pattern. Compared to the unmixed case, flavor mixing weakens the exclusion limits in the $ee$ and $\mu\mu$ channels. Under maximal mixing, the $e\mu$ channel is the most sensitive individual channel, while under PMNS‑like mixing the $ee$ channel provides the strongest sensitivity. Combining all dilepton channels further improves the reach, especially in the maximal‑mixing scenario.

At LHC Run 2, the combined $\ell\ell$ channel already provides significant constraints on the parameter space. The HL‑LHC, with its increased luminosity, substantially extends the accessible mass range. Under maximal (PMNS‑like) mixing, the HL‑LHC can exclude $m_{W_R}$ up to $6.7$ ($6.3$)~TeV for $m_{N_R}=2$~TeV and $m_{N_R}$ up to $4.4$ ($4.1$)~TeV for $m_{W_R}=5.6$ ($5.4$)~TeV. These limits are substantially stronger than those obtained at Run 2.

We have also explored generic mixing configurations in the $|V_{e1}|\text{--}|V_{\mu1}|$ plane. At the LHC Run 2 benchmark point $(m_{W_R}, m_{N_R}) = (5~\text{TeV}, 3~\text{TeV})$, the expected sensitivity of the $ee$, $e\mu$, and $\mu\mu$ channels decreases as $|V_{\tau1}|$ increases, leading to an exclusion boundary approximately characterized by $\sqrt{|V_{e1}|^2 + |V_{\mu1}|^2} \sim 0.8\text{--}0.85$ for the SS+OS dilepton analysis. At the HL-LHC, the expected sensitivity is significantly extended, probing down to $\sqrt{|V_{e1}|^2 + |V_{\mu1}|^2} \sim 0.7\text{--}0.8$ for heavier benchmark masses.

Finally, we investigated complementary constraints from low-energy CLFV observables. We consider a benchmark scenario with degenerate charged scalars ($m_{\Delta_{L,R}^{--}} = m_{\Delta_L^-} = m_{W_R}$) and a relative mass splitting $(m_{N_2}-m_{N_1})/m_{N_1}=0.01$. Under these assumptions, current CLFV bounds, when combined with LHC Run~2 searches, exclude $m_{W_R} \lesssim 6~\text{TeV}$ for any $m_{N_R}$. Future CLFV experiments will extend this reach, probing $m_{W_R} \gtrsim 7~\text{TeV}$ and potentially exceeding the HL-LHC sensitivity. These results demonstrate the strong complementarity between direct collider searches and indirect CLFV probes in testing the LRSM parameter space.

In summary, our results highlight the essential role of right-handed lepton flavor mixing in interpreting collider searches for right-handed gauge bosons and neutrinos. The combination of machine learning techniques with
an explicit treatment of $e,\mu$ flavor effects provides 
an effective strategy
for probing TeV-scale new physics at current and future experiments.

\section{Acknowledgments}

We would like to thank Frank Deppisch, Bingxuan Liu, Michihisa Takeuchi, and Lailin Xu for helpful discussions and comments.
This work is supported by the National Natural Science Foundation of China under Grants No.~12347105 and No.~12505127, and the Guangdong Basic and Applied Basic Research Foundation (2024A1515012668).

\appendix
\counterwithin{figure}{section}
\counterwithin{table}{section}

%%%%%%%%%%%%%%%%%%%%%%%%%%%%%%%%%%%%%%%%%
\section{More details of ML analysis}
\label{app:ML_OS}
%%%%%%%%%%%%%%%%%%%%%%%%%%%%%%%%%%%%%%%%%

\begin{figure*}[!b]
    \centering
    \includegraphics[width=0.7\linewidth]{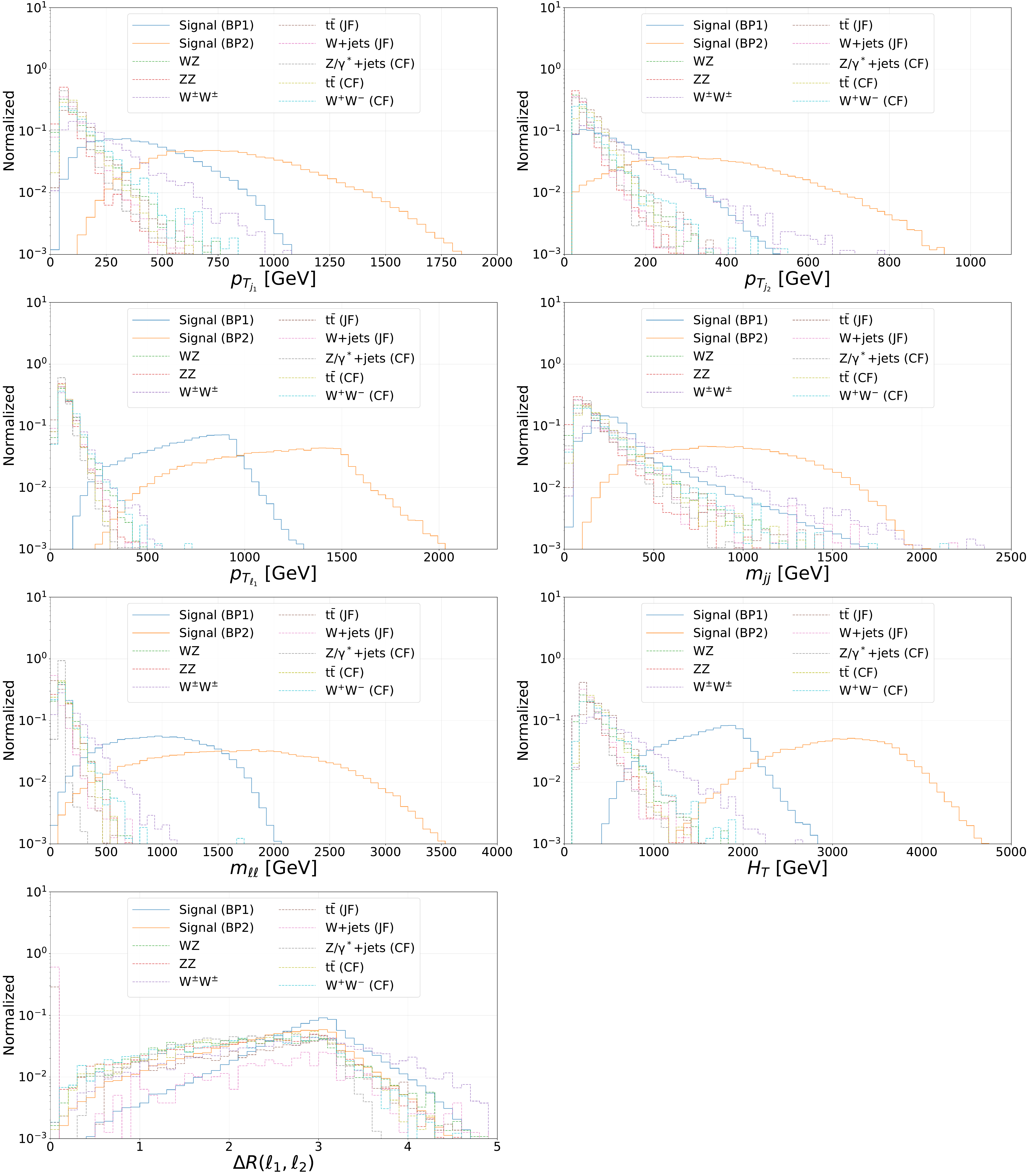}
  \caption{Normalized distributions for SS dilepton signals and SM backgrounds at the LHC with $\sqrt{s} = 13~\text{TeV}$. Signals correspond to $(m_{W_R}, m_{N_R}) = (2~\text{TeV},\, 0.5~\text{TeV})$ (blue solid) and $(4~\text{TeV},\, 2~\text{TeV})$ (orange solid). Left panels (top to bottom): $p_{T_{j_1}}$, $p_{T_{\ell_1}}$, $m_{jj}$, $\Delta R (\ell_1,\ell_2)$. Right panels (top to bottom): $p_{T_{j_2}}$, $m_{\ell\ell}$, $H_T$.}
  \label{fig:distribution}
\end{figure*}

The normalized distributions of the input variables for the SS dilepton signal and the corresponding SM backgrounds are shown in Fig.~\ref{fig:distribution}. 
Signal events populate higher-value regions of several observables compared to backgrounds, a pattern that the DNN captures during training. This is especially pronounced for the higher-mass benchmark, providing correlated information that enhances discrimination.

\begin{figure}[!htbp]
  \centering
  \includegraphics[width=0.23\textwidth]{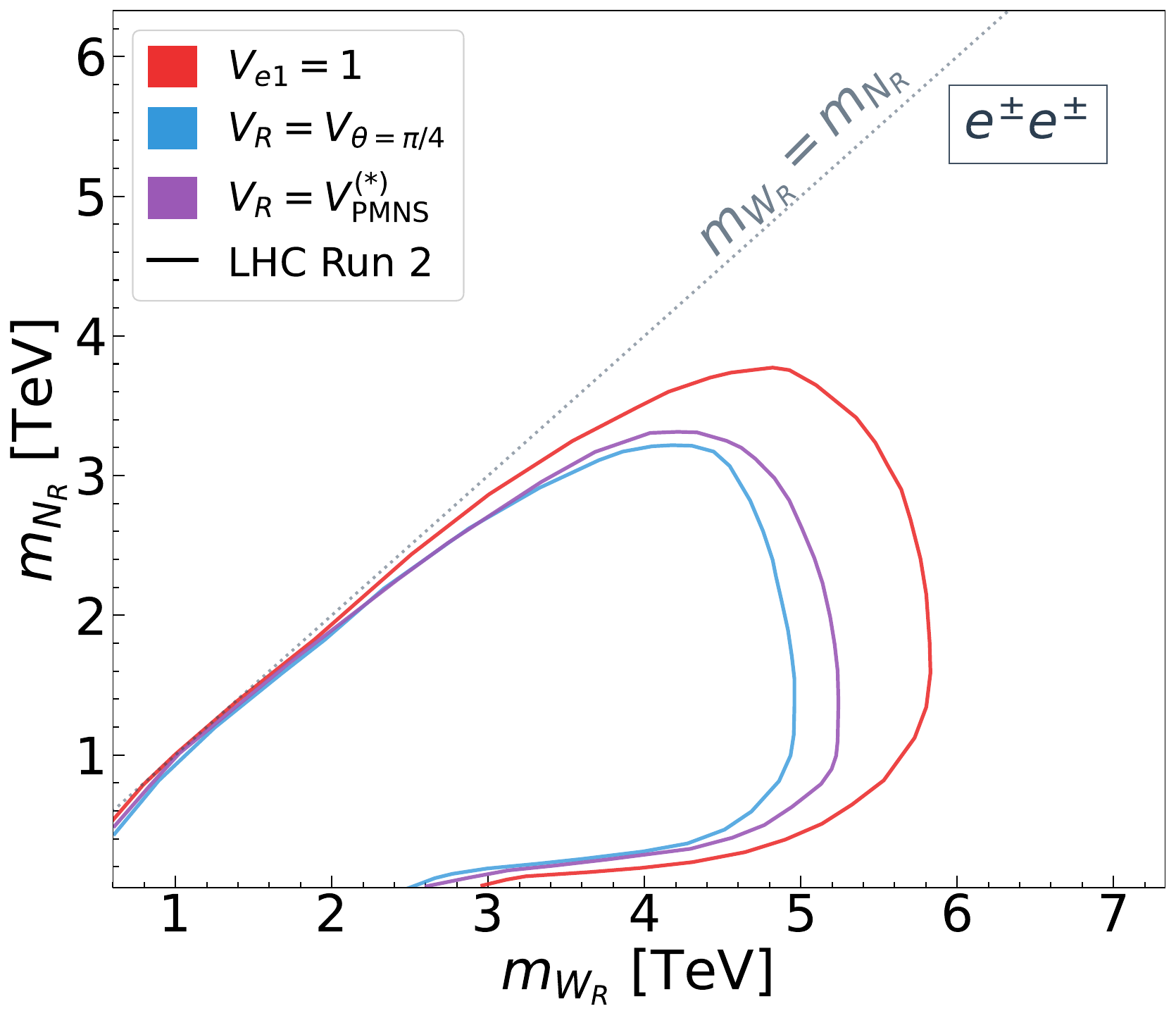}
  \hspace{0.1cm}
  \includegraphics[width=0.23\textwidth]{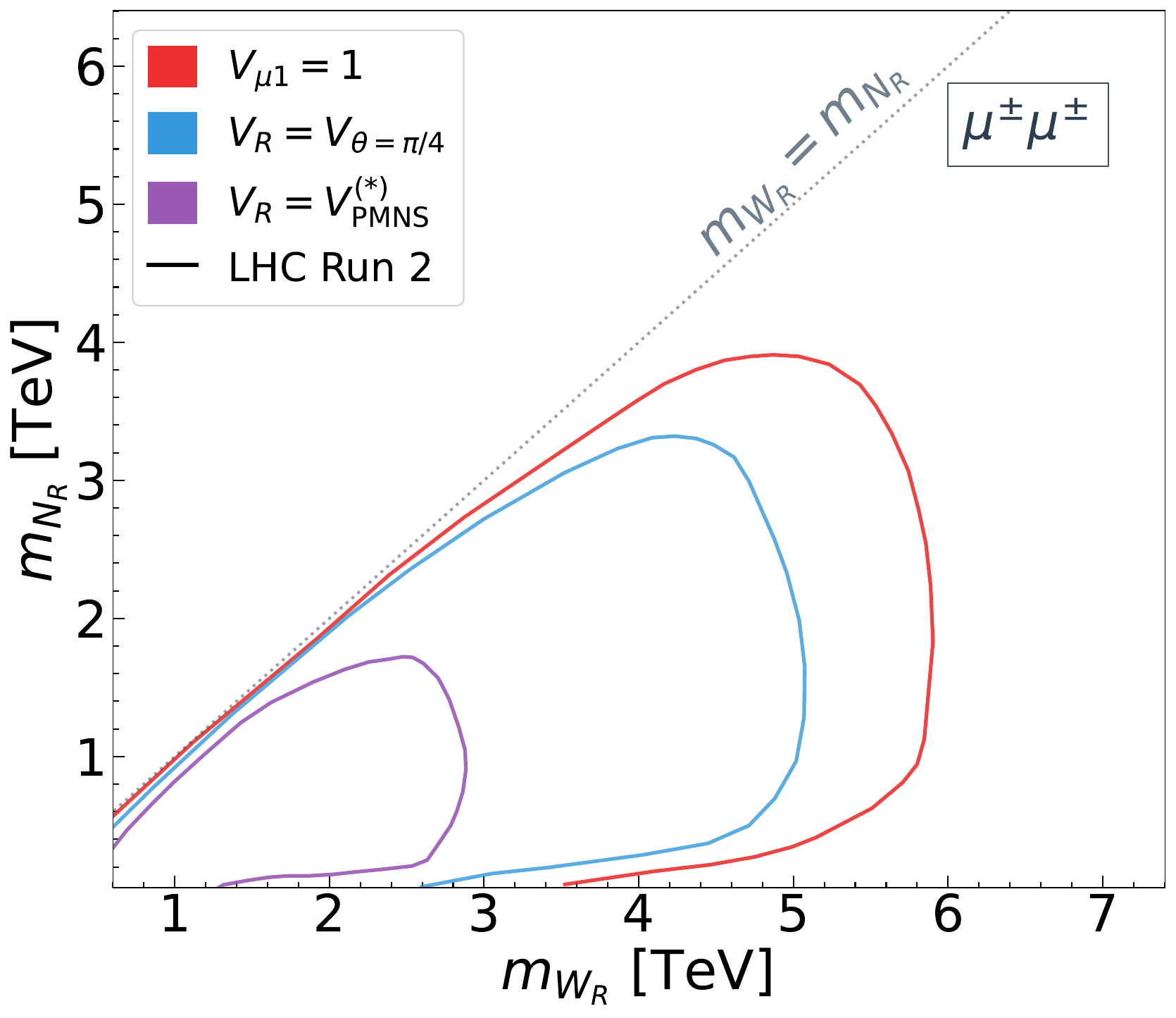}\\[0.1cm]
  \includegraphics[width=0.23\textwidth]{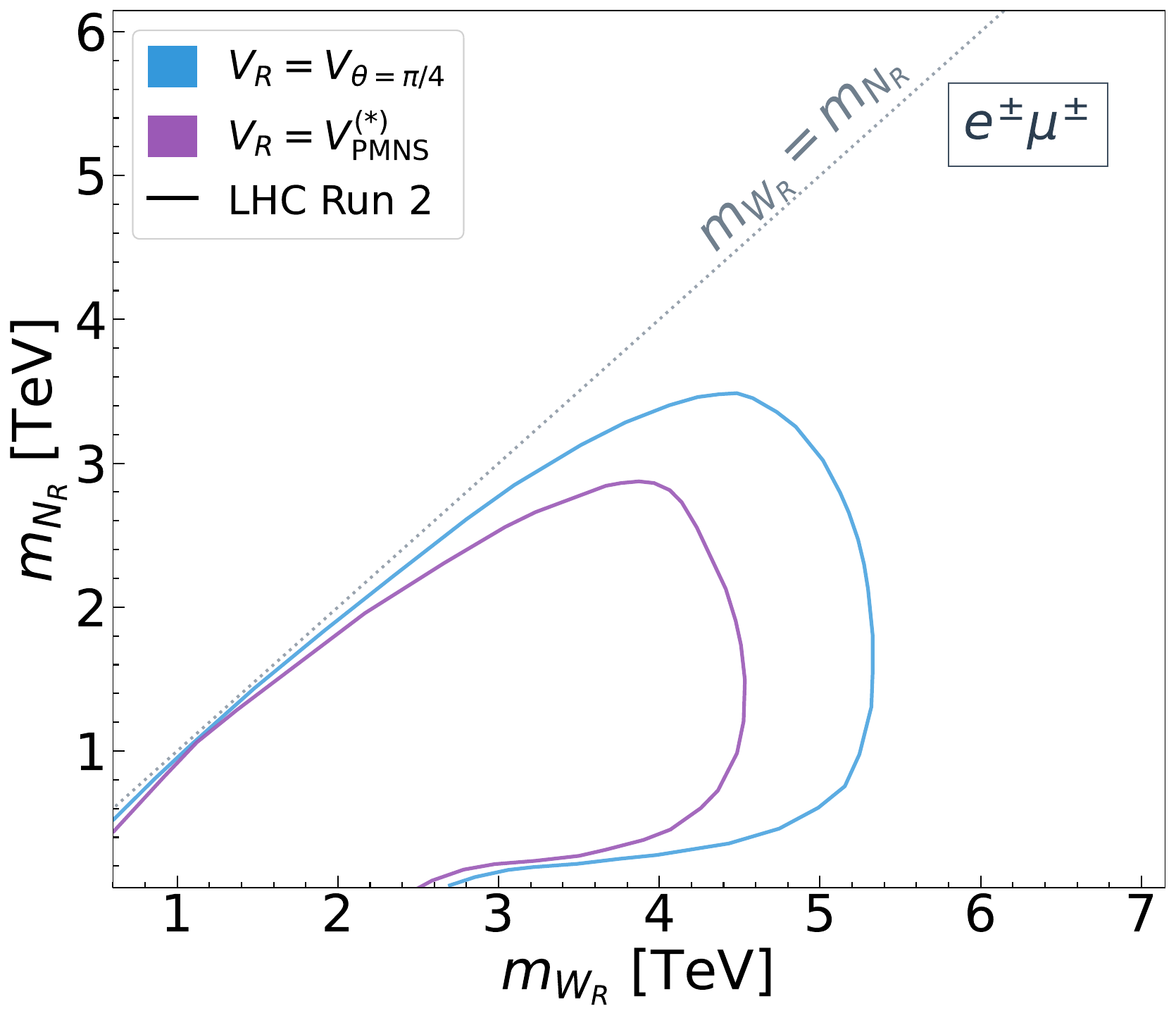}
  \hspace{0.1cm}
  \includegraphics[width=0.23\textwidth]{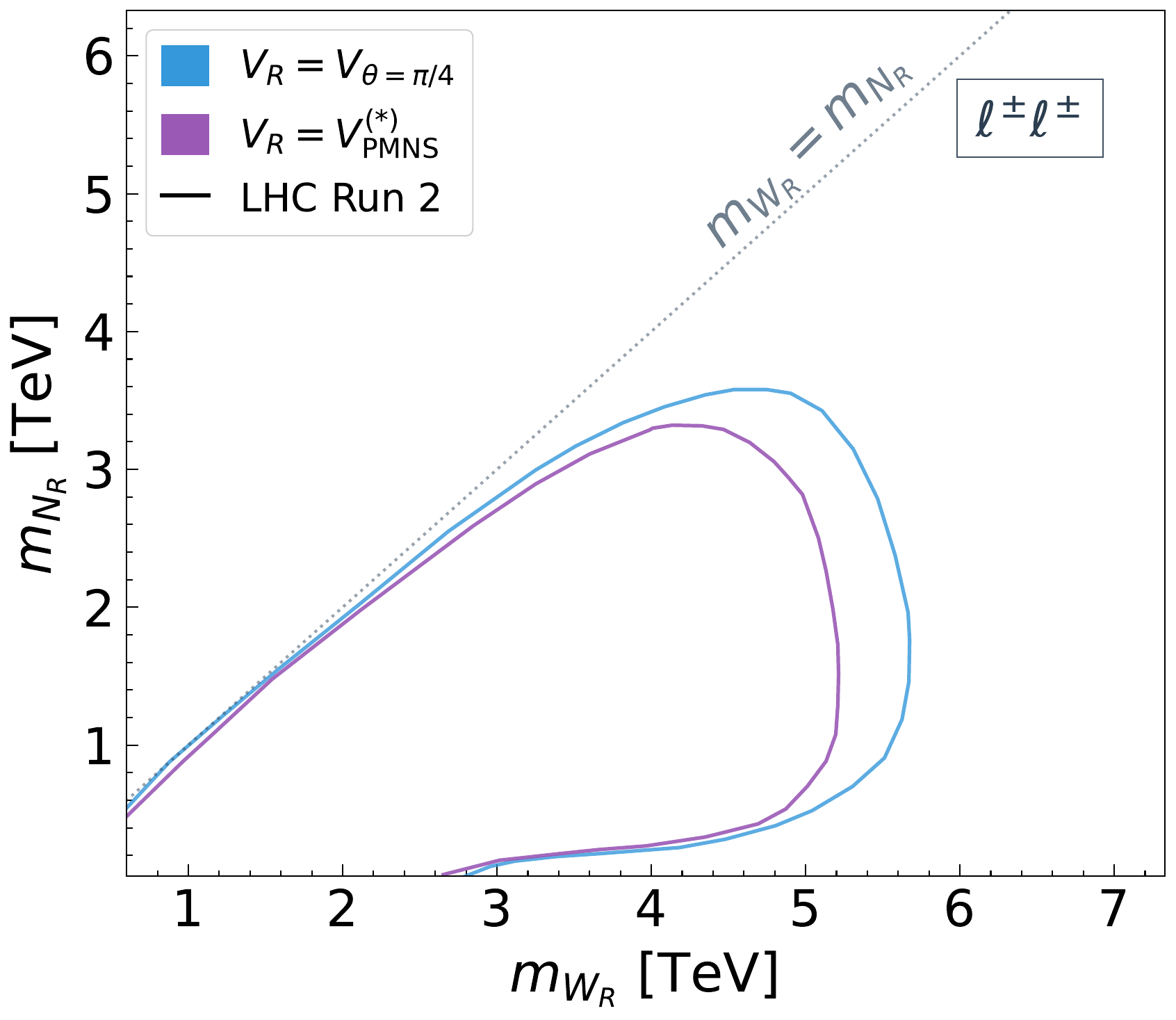}
  \caption{95\% C.L. constraints on $m_{W_R}$ and $m_{N_R}$ from SS dilepton searches at LHC Run 2. The panels correspond to the $e^\pm e^\pm$ (upper left), $\mu^\pm \mu^\pm$ (upper right), $e^\pm \mu^\pm$ (lower left), and combined $\ell^\pm \ell^\pm$ ($\ell=e,\mu$) (lower right) channels. The blue and purple contours represent the maximal-mixing scenario ($V_R=V_{\theta=\pi/4}$) and the PMNS-like scenario ($V_R=V_{\mathrm{PMNS}}^{(*)}$), respectively. In the $e^\pm e^\pm$ and $\mu^\pm \mu^\pm$ channels, the red contours denote the unmixed cases ($V_{e1}=1$ or $V_{\mu1}=1$). The gray dotted line in each panel marks the kinematic threshold $m_{W_R}=m_{N_R}$.}
  \label{fig:mass_reach_SS}
\end{figure}

\begin{figure}[!htbp]
  \centering
  \includegraphics[width=0.25\textwidth]{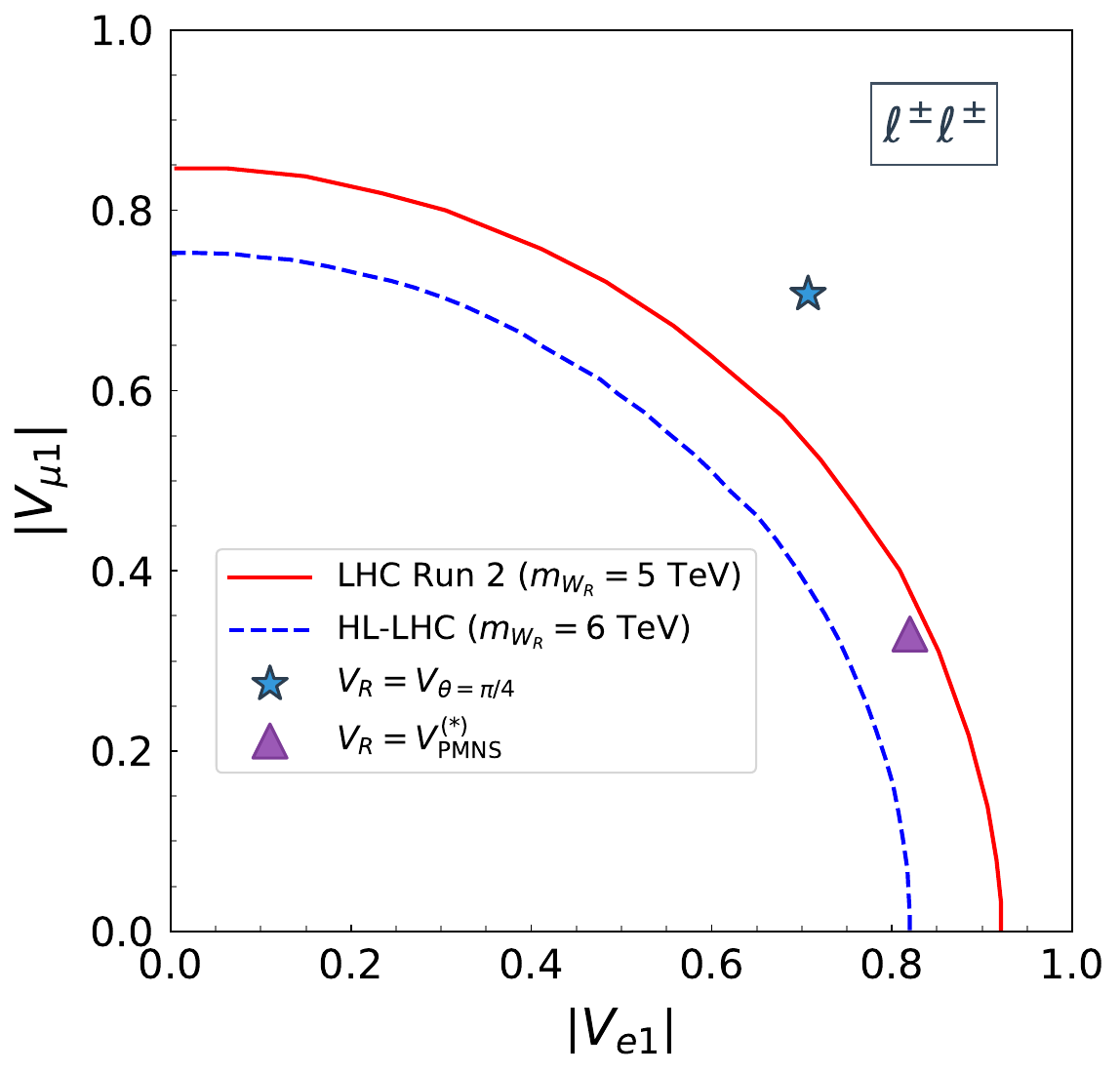}
  \caption{95\% C.L. constraints on $|V_{e1}|$ and $|V_{\mu1}|$ with the benchmark $m_{N_R}=3$~TeV in the SS dilepton channels. The red solid contour shows the limit from LHC Run 2 for $m_{W_R}=5$~TeV, while the blue dashed contour indicates the HL-LHC sensitivity for $m_{W_R}=6$~TeV. The star marks the maximal-mixing scenario, and the triangle marks the PMNS-like scenario.}
  \label{fig:mixing_bound_SS}
\end{figure}

The expected exclusion limits on $m_{W_R}$ and $m_{N_R}$ in the SS dilepton searches at LHC Run 2 are shown in Fig.~\ref{fig:mass_reach_SS}. The sensitivity exhibits the same dependence on mixing scenarios as in the combined SS+OS analysis, as discussed in Section~\ref{sec:exclusion_mass}.
Figure~\ref{fig:mixing_bound_SS} shows the constraints in the $|V_{e1}|\text{--}|V_{\mu1}|$ plane for the SS dilepton channel. The exclusion boundary is characterized by $\sqrt{|V_{e1}|^2 + |V_{\mu1}|^2} \sim 0.85\text{--}0.9$ at LHC Run 2 and $\sim 0.75\text{--}0.8$ at the HL-LHC. With only SS searches, the maximal-mixing scenario is excluded at both LHC Run 2 and the HL-LHC, whereas the PMNS-like scenario is not excluded by Run 2 but would require HL-LHC sensitivity. This differs from the combined SS+OS analysis (cf. Fig.~\ref{fig:mixing_bound_SS+OS}).

The OS dilepton channel shares the same signal topology and kinematic features as the SS channel, so the same set of input variables (cf. Eq.~\eqref{eq:inputs}) is used for the DNN classifier. As in the SS analysis, two benchmark models (BP1 and BP2) are employed to ensure stable performance across the parameter space. However, the background composition differs: the $t\bar{t}$ and $Z/\gamma^*+\text{jets}$ processes contribute directly in the OS channel, leading to a substantially larger number of background events after the baseline selection.

\begin{figure*}[!t]
  \centering
  \includegraphics[width=0.4\textwidth]{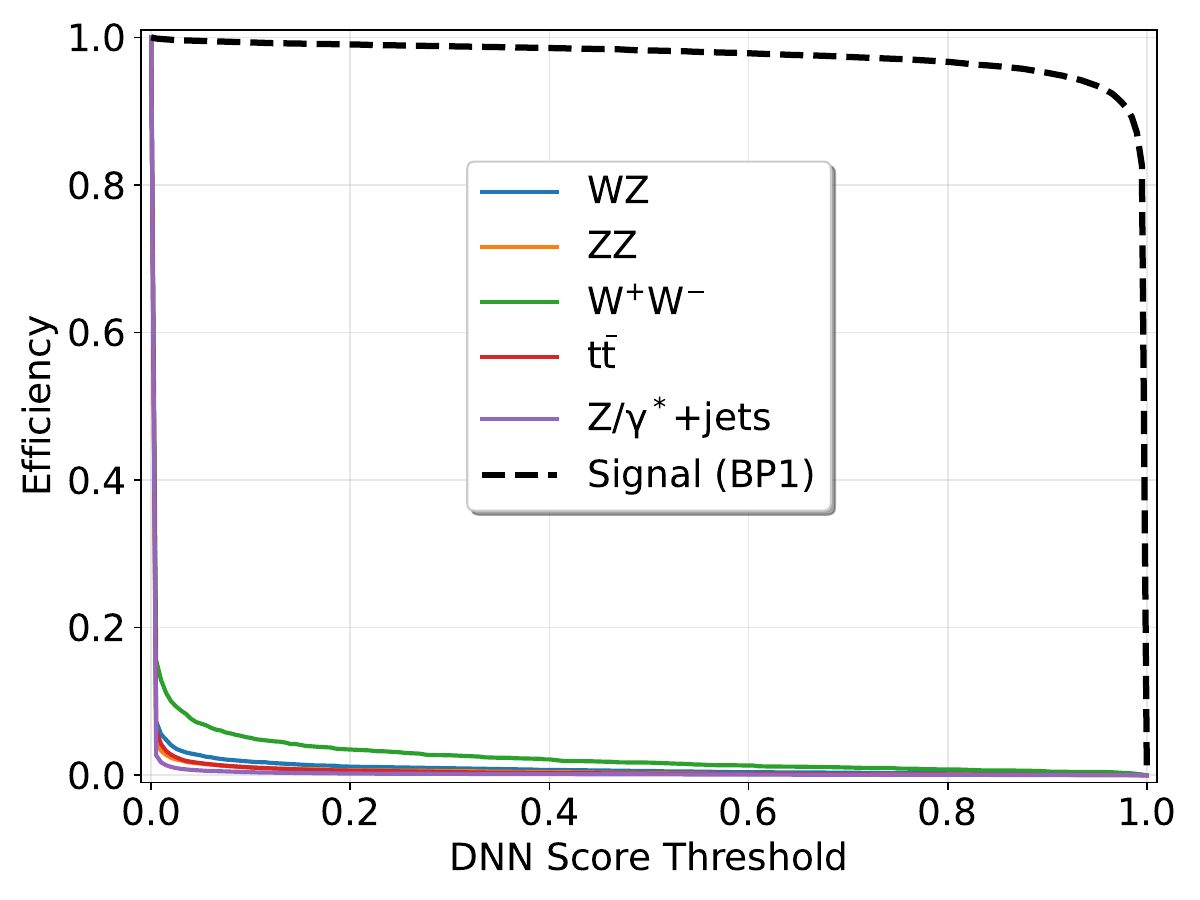}
  \hspace{0.3cm}
  \includegraphics[width=0.4\textwidth]{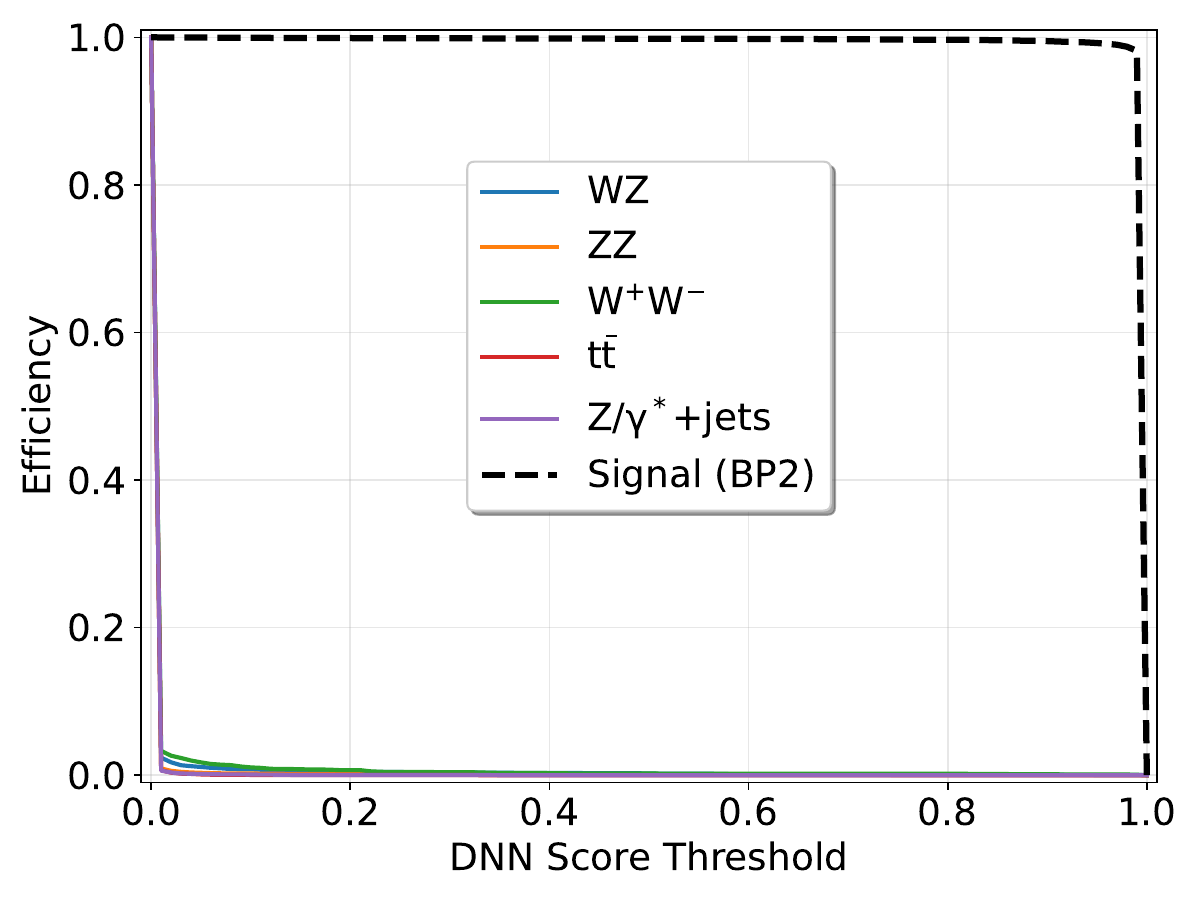}
  \caption{
 Same as Fig.~\ref{fig:DNN-eff_SS}, but for the OS dilepton searches.
  }
  \label{fig:DNN-eff_OS}
\end{figure*}

Figure~\ref{fig:DNN-eff_OS} shows the selection efficiencies as a function of the DNN score threshold for the low‑mass and high‑mass models in the OS dilepton searches. For the low‑mass benchmark, the DNN achieves strong background rejection while maintaining high signal efficiency. A similar analysis is performed for the high‑mass benchmark.

\begin{table}[!b]
\small
\setlength{\tabcolsep}{3pt}
\caption{Cross sections for the OS dilepton signal and background processes for the low-mass model. Values are shown at three stages: after the basic cuts ($\sigma_{\text{basic}}$), after the baseline selection ($\sigma_{\ell\ell jj}$), and after the optimal DNN score selection ($\sigma_{\text{DNN}}$). The bottom rows provide the comparison of the total background and signal cross sections.}
\label{tab:OS_cutcross_low}
\begin{ruledtabular}
\begin{tabular}{lccc}
Process & $\sigma_{\text{basic}}$ (pb) & $\sigma_{\ell\ell jj}$ (pb) & $\sigma_{\text{DNN}}$ (pb) \\
\hline
$WZ$ & $4.31\!\times\!10^{-1}$ & $4.49\!\times\!10^{-2}$ & $3.59\!\times\!10^{-5}$ \\
$W^+W^-$ & $3.73\!\times\!10^{-1}$ & $1.01\!\times\!10^{-1}$ & $1.92\!\times\!10^{-4}$ \\
$ZZ$ & $4.15\!\times\!10^{-2}$ & $4.34\!\times\!10^{-3}$ & $8.68\!\times\!10^{-7}$ \\
$t\bar{t}$ & $1.81\!\times\!10^{1}$ & $4.99$ & $1.75\!\times\!10^{-5}$ \\
$Z/\gamma^*+\text{jets}$ & $5.33\!\times\!10^{1}$ & $2.55\!\times\!10^{1}$ & $1.39\!\times\!10^{-4}$ \\
\hline
Total bkg & $7.22\!\times\!10^{1}$ & $3.06\!\times\!10^{1}$ & $3.85\!\times\!10^{-4}$ \\
Signal & $5.21\!\times\!10^{-2}$ & $2.65\!\times\!10^{-2}$ & $2.31\!\times\!10^{-2}$ \\
\end{tabular}
\end{ruledtabular}
\end{table}

\begin{table}[!t]
\small
\setlength{\tabcolsep}{3pt}
\caption{Event yields after the optimal DNN score selection for the low-mass model. The backgrounds (individual categories and total) and the OS dilepton signal for BP1 are shown for the three OS dilepton channels. A dash (--) indicates that the corresponding background contribution is negligible. The last row gives the signal efficiency in each channel.}
\label{tab:OS_cutevent_low}
\centering
\begin{tabular}{lccc}
\hline
Process & $e^\pm \mu^\mp$ & $e^+ e^-$ & $\mu^+ \mu^-$ \\
\hline
Multiboson & 10 & 12 & 10 \\
$t\bar{t}$ & 1 & -- & 1 \\
$Z/\gamma^* + \text{jets}$ & -- & 10 & 10 \\
\hline
Total bkg & 11 & 22 & 21 \\
Signal & 1611 & 617 & 983 \\
Eff [\%] & 44 & 34 & 54 \\
\hline
\end{tabular}
\end{table}

\begin{table}[!t]
\small
\setlength{\tabcolsep}{3pt}
\caption{Same as Table~\ref{tab:OS_cutcross_low}, but for the high-mass model.}
\label{tab:OS_cutcross_high}
\begin{ruledtabular}
\begin{tabular}{lccc}
Process & $\sigma_{\text{basic}}$ (pb) & $\sigma_{\ell\ell jj}$ (pb) & $\sigma_{\text{DNN}}$ (pb) \\
\hline
$WZ$ & $4.31\!\times\!10^{-1}$ & $4.49\!\times\!10^{-2}$ & $5.23\!\times\!10^{-6}$ \\
$W^+W^-$ & $3.73\!\times\!10^{-1}$ & $1.01\!\times\!10^{-1}$ & $1.05\!\times\!10^{-5}$ \\
$ZZ$ & $4.15\!\times\!10^{-2}$ & $4.34\!\times\!10^{-3}$ & $\lesssim 10^{-7}$ \\
$t\bar{t}$ & $1.81\!\times\!10^{1}$ & $4.99$ & $4.62\!\times\!10^{-6}$ \\
$Z/\gamma^*+\text{jets}$ & $5.33\!\times\!10^{1}$ & $2.55\!\times\!10^{1}$ & $2.09\!\times\!10^{-5}$ \\
\hline
Total bkg & $7.22\!\times\!10^{1}$ & $3.06\!\times\!10^{1}$ & $4.13\!\times\!10^{-5}$ \\
Signal & $4.16\!\times\!10^{-4}$ & $2.27\!\times\!10^{-4}$ & $2.20\!\times\!10^{-4}$ \\
\end{tabular}
\end{ruledtabular}
\end{table}

\begin{table}[H]
\small
\setlength{\tabcolsep}{3pt}
\caption{Same as Table~\ref{tab:OS_cutevent_low}, but for the high-mass model.}
\label{tab:OS_cutevent_high}
\centering
\begin{tabular}{lccc}
\hline
Process & $e^\pm \mu^\pm$ & $e^\pm e^\pm$ & $\mu^\pm \mu^\pm$ \\
\hline
Multiboson & 0.89 & 0.67 & 0.62 \\
$t\bar{t}$ & 0.23 & 0.16 & 0.25 \\
$Z/\gamma^* + \text{jets}$ & -- & 1.36 & 1.55 \\
\hline
Total bkg & 1.12 & 2.19 & 2.42 \\
Signal & 14.84 & 6.52 & 9.24 \\
Eff [\%] & 50.31 & 44.15 & 62.69 \\
\hline
\end{tabular}
\end{table}

By maximizing the expected significance $Z$, we obtain optimal DNN score thresholds of 0.99 for both the low-mass and high-mass models in the OS dilepton searches.
Table~\ref{tab:OS_cutcross_low} presents the cross sections after each selection stage for the OS dilepton channel. As in the SS channel, signal efficiency remains high while backgrounds are strongly suppressed. Despite larger contributions from $t\bar{t}$ and $Z/\gamma^*+\text{jets}$, the DNN reduces the total background by nearly five orders of magnitude, making the OS background comparable to the SS after the selections.

The event yields after the DNN selection, expected at LHC Run 2, are summarized in Table~\ref{tab:OS_cutevent_low}. The corresponding efficiencies are $\epsilon_{e\mu}=44\%$, $\epsilon_{ee}=34\%$, and $\epsilon_{\mu\mu}=54\%$. This pattern is consistent with the SS case, where better muon reconstruction leads to higher efficiency in the $\mu^+\mu^-$ channel.

From Tables~\ref{tab:OS_cutcross_high} and~\ref{tab:OS_cutevent_high} for the high‑mass model, background suppression is more efficient than in the low‑mass model with the optimal DNN threshold of 0.99. Despite larger backgrounds than in the SS channel, the DNN reduces the OS background to a level comparable to the SS channel after the selections. For an integrated luminosity of $139~\text{fb}^{-1}$, some expected event yields are below unity. Following the convention of Table~\ref{tab:cutevent_high} in Section~\ref{sec:ML-analysis}, we show all yields with two decimal places for uniform precision.

%%%%%%%%%%%%%%%%%%%%%%%%%%%%%%%%%%%%%%%%%
\section{DNN implementation and validation}
\label{app:DNN}
%%%%%%%%%%%%%%%%%%%%%%%%%%%%%%%%%%%%%%%%%

The DNN classifiers are implemented in \texttt{PyTorch Lightning}~\cite{falcon2020pytorch} and take the kinematic variables defined in Eq.~\eqref{eq:inputs} as inputs. The input variables are standardized using the \texttt{StandardScaler} routine from \texttt{scikit-learn}~\cite{scikit-learn} before being passed to the network. The network consists of two fully connected hidden layers with 64 nodes each and ReLU activation functions. The output layer contains two nodes, corresponding to the background and signal classes. The signal score is obtained from the softmax probability of the signal class. Training is performed with the Adam optimizer~\cite{Kingma:2014vow}, with an initial learning rate of $10^{-3}$, a batch size of 1024, and a cross-entropy loss. The network is trained for 50 epochs. The simulated samples are split into training, validation, and test samples with fractions of 80\%, 10\%, and 10\%, respectively. 
No event weights are applied in the DNN training loss; all simulated events entering the training samples are treated with equal weight. The process cross sections are not used during training, but are applied after the DNN selection to convert the signal and background efficiencies into expected yields.
The training and validation loss and accuracy are monitored throughout the training. Both the training and validation losses decrease steadily at the beginning of the training and subsequently reach stable plateaus, while the corresponding accuracies show consistent convergence. No increase of the validation loss after convergence or divergence between the training and validation performance is observed, indicating no evident sign of overtraining.

\begin{figure}[!htbp]
\centering
\includegraphics[width=0.22\textwidth]{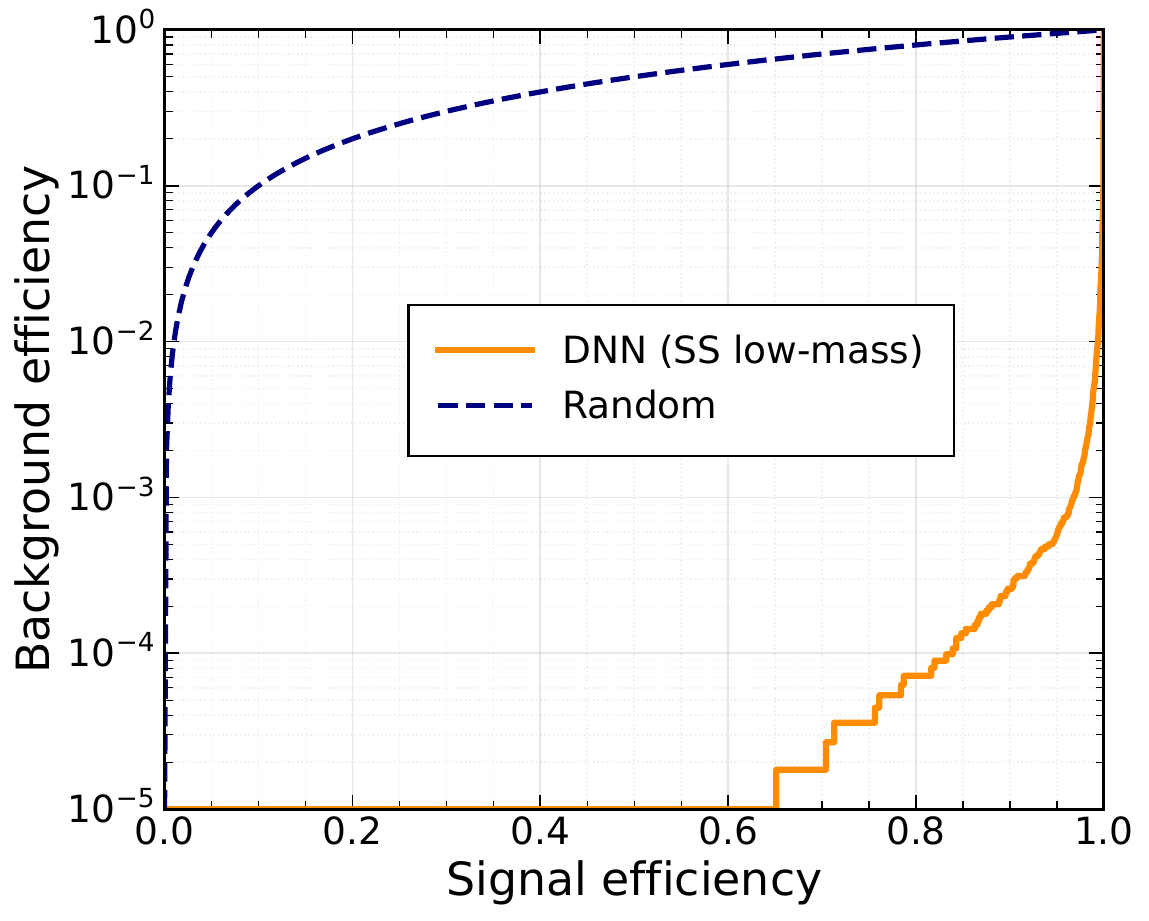}
\includegraphics[width=0.22\textwidth]{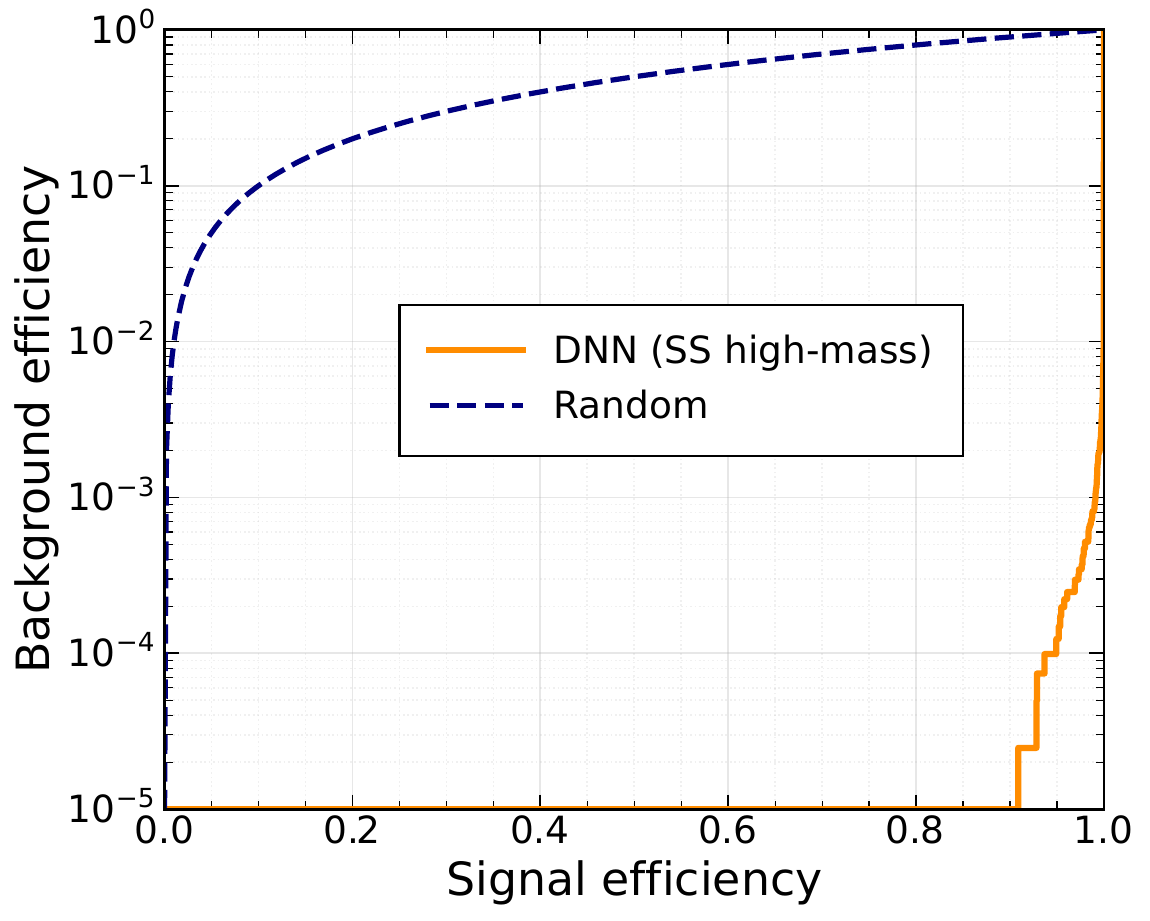}
\caption{ROC curves for the low-mass and high-mass DNN classifiers used in the SS dilepton analysis, shown in the left and right panels, respectively.}
\label{fig:ROC_curves}
\end{figure}

To validate the performance of the DNN classifiers used in the analysis, we show the receiver operating characteristic (ROC) curves for the SS dilepton channel in Fig.~\ref{fig:ROC_curves}. Two independent DNN classifiers are trained using BP1 and BP2, corresponding to the low-mass and high-mass models, respectively.

The ROC curves display the background efficiency versus the signal efficiency, with the background efficiency plotted on a logarithmic scale to highlight the rejection power in the high-purity (tight DNN-score) regime targeted by the final event selection. For both the low-mass and high-mass classifiers, the resulting curves are clearly separated from the random-classifier baseline, demonstrating efficient discrimination between signal and background.

\begin{figure}[!htbp]
\centering
\includegraphics[width=0.22\textwidth]{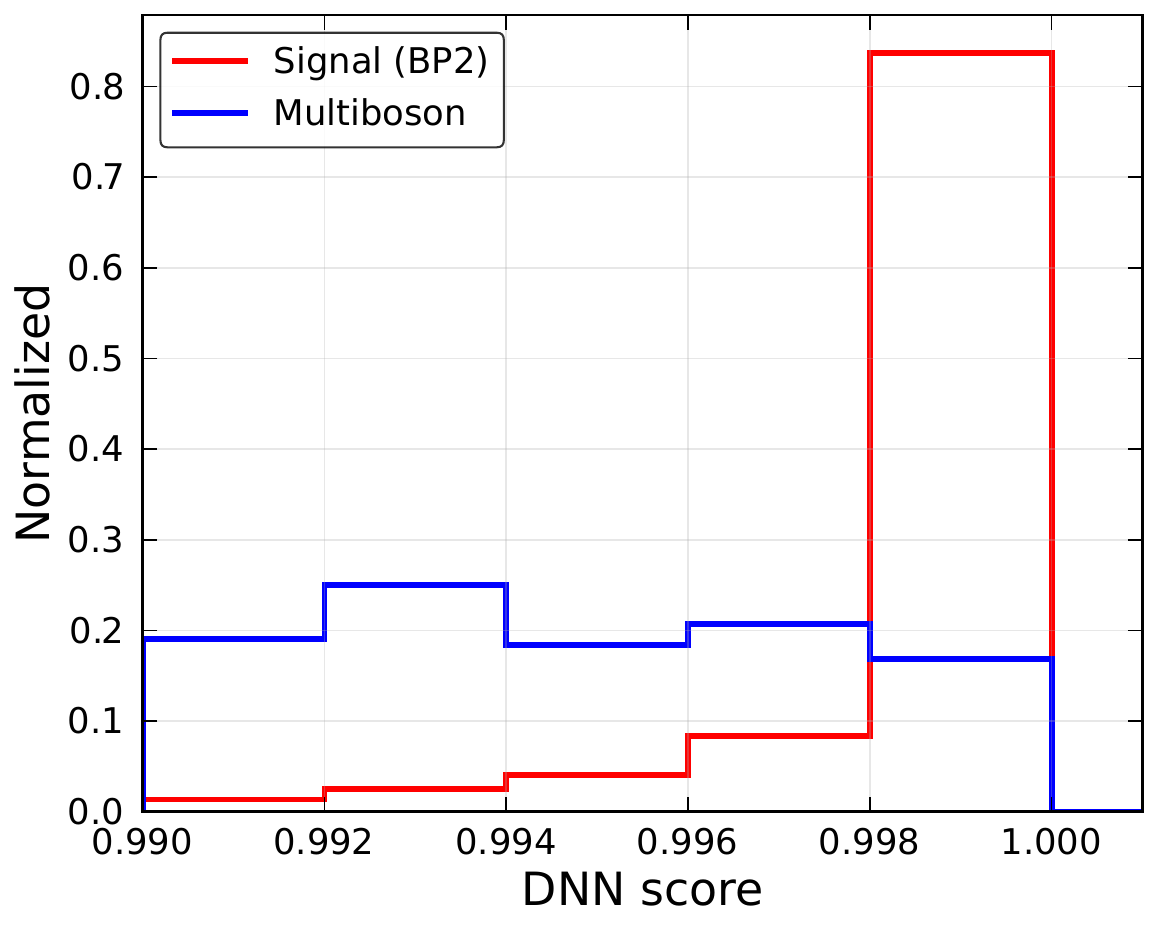}
\includegraphics[width=0.22\textwidth]{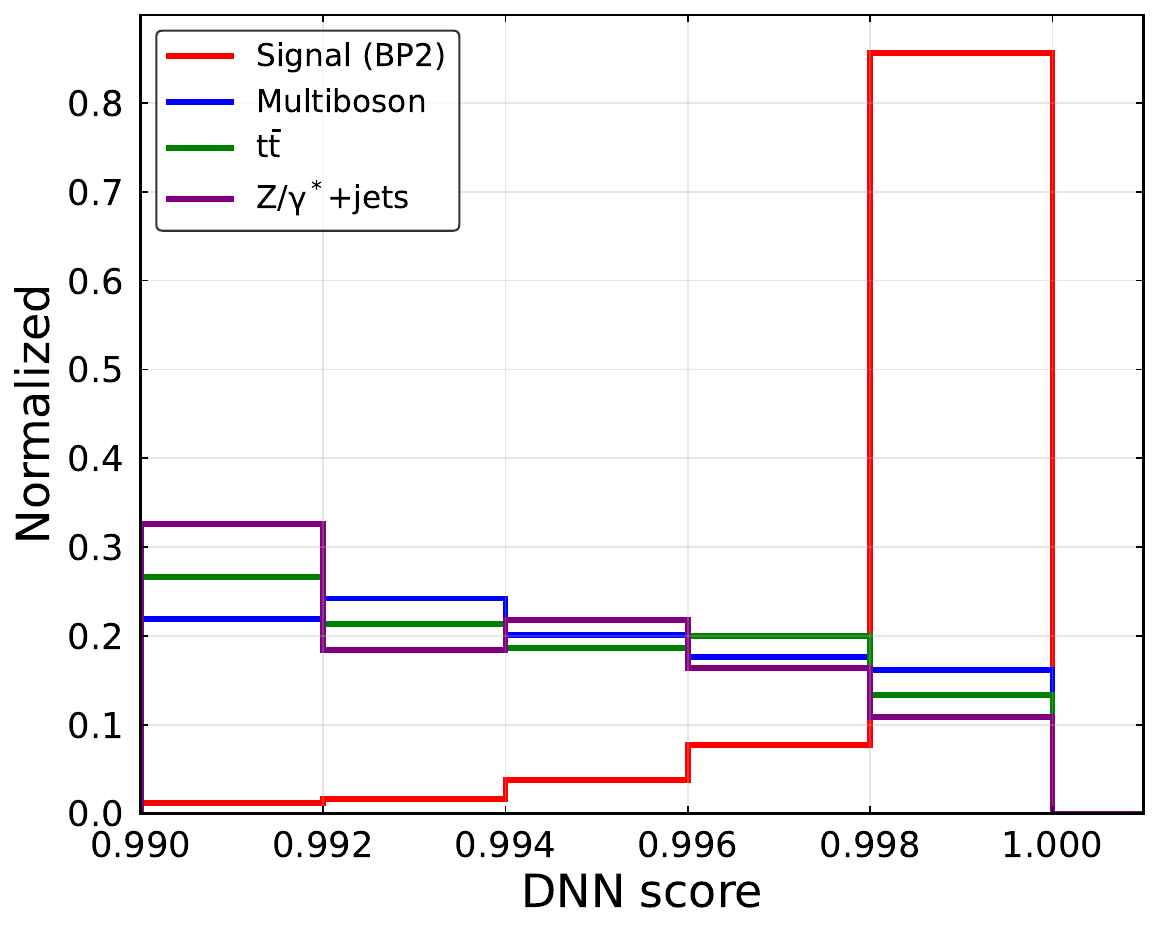}
\caption{Normalized DNN-score distributions in the high-score region after applying $D>0.99$ for the high-mass model. The left and right panels show the SS and OS dilepton analyses, respectively, including the signal and the corresponding SM backgrounds.}
\label{fig:DNN_tails}
\end{figure}

The OS dilepton channel is treated analogously. As an additional check of the DNN behavior in the high-score region, Fig.~\ref{fig:DNN_tails} shows the normalized DNN-score distributions in the tail region $D>0.99$ for the high-mass classifiers used in the SS and OS dilepton analyses. 
In the SS channel, $\mathcal{O}(10^6)$ unweighted simulated multiboson events pass the baseline selection, with $\mathcal{O}(10^2)$--$\mathcal{O}(10^3)$ events remaining after the DNN selection. The JF and CF samples contain $\mathcal{O}(10^6)$ and $\mathcal{O}(10^7)$ unweighted events after the baseline selection, respectively, but are reduced to only $\mathcal{O}(10)$--$\mathcal{O}(10^3)$ events or fewer after the tight DNN requirement, depending on the specific background\,\footnote{For the low-mass classifier, the multiboson, JF, and CF backgrounds are comparable after the DNN selection (cf. Table~\ref{tab:cutcross_low}). At least $\mathcal{O}(10^2)$ unweighted events remain for each background component in our simulation, indicating that the MC statistical uncertainty in the tail region is under control.}.

After rescaling to the expected cross sections, the JF and CF contributions are at least one order of magnitude smaller than the multiboson backgrounds (cf. Table~\ref{tab:cutcross_high}), even after accounting for their modeling uncertainties, taken as 10\% and 17\%~\cite{Curtin:2013zua,Muskinja:2018eba}, respectively. Therefore, the JF and CF backgrounds are not shown in the DNN-score tail distributions in Fig.~\ref{fig:DNN_tails} and are not included in Table~\ref{tab:cutevent_high}.
In contrast, for the OS channel, the DNN-score tail distributions of the three dominant backgrounds are all shown in Fig.~\ref{fig:DNN_tails}. 
After the DNN selection, up to $\mathcal{O}(10^2)$ unweighted events for each background component in our simulation are retained, which keeps the MC statistical uncertainties in the tail region under control.
These validation checks support the robustness of the DNN-based event selection.

\begin{table}[htbp]
\centering
\caption{Validation of the trained DNN classifiers at representative mass points. The asterisks indicate the training benchmark points BP1 and BP2.}
\begin{tabular}{c c c c c}
\hline
Region & $m_{W_R}$ (TeV) & $m_{N_R}$ (TeV) & $m_{N_R}/m_{W_R}$ 
& $\epsilon_{\rm DNN}$ \\
\hline
low$^*$        & 2 & 0.5 & 0.250 & 94.8\% \\
low        & 2 & 1.0 & 0.500 & 92.1\% \\
low        & 2 & 1.5 & 0.750 & 78.7\% \\
high       & 4 & 0.5 & 0.125 & 92.3\% \\
high       & 4 & 1.0 & 0.250 & 95.4\% \\
high$^*$       & 4 & 2.0 & 0.500 & 96.9\% \\
high       & 4 & 3.0 & 0.750 & 94.8\% \\
high       & 4 & 3.5 & 0.875 & 82.3\% \\
\hline
\end{tabular}
\label{tab:dnn_generalization}
\end{table}

For each signal region, the fixed trained DNN classifier is applied to all scanned mass points using the same preprocessing and DNN-score requirement. The signal efficiency is evaluated separately at each mass point and is then used to compute the expected yield and significance, and derive the expected exclusion limits.

To assess the applicability of the trained classifiers in different regions, we examine the benchmark points together with representative non-training mass points covering different values of $m_{N_R}/m_{W_R}$. Table~\ref{tab:dnn_generalization} shows the corresponding DNN cut efficiencies,  $\epsilon_{\rm DNN} \equiv \sigma_{\rm DNN}/\sigma_{\ell\ell jj}$.
Here ``low'' and ``high'' indicate that the signal point is evaluated using the low-mass and high-mass classifiers trained using BP1 and BP2, respectively, as defined in Section~\ref{sec:ML-analysis}. The asterisks indicate the training
benchmark points.

As shown in Table~\ref{tab:dnn_generalization}, the DNN cut efficiencies remain high for both the benchmark points and the additional non-training mass points. The efficiencies remain stable over the tested regions, with only an expected moderate decrease at larger values of $m_{N_R}/m_{W_R}$, where the $N_R$ and $W_R$ masses become closer. Nevertheless, the signal still retains characteristic high-energy kinematic features, allowing it to be effectively distinguished from the background.

This validation shows that the trained classifiers provide stable signal selection away from the benchmark points. While the exclusion contours could be further improved by training classifiers at multiple values of $m_{N_R}/m_{W_R}$, the stability check presented here suggests that such a finer training strategy would not qualitatively change our conclusions.

%%%%%%%%%%%%%%%%%%%%%%%%%%%%%%%%%%%%%%%%%
\section{CLFV processes: analytic expressions}
\label{app:LFV}
%%%%%%%%%%%%%%%%%%%%%%%%%%%%%%%%%%%%%%%%%

\begin{figure}[!htbp]
\centering
\includegraphics[width=0.22\textwidth]{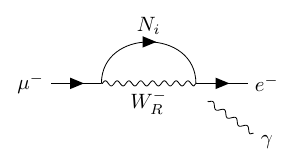}
\includegraphics[width=0.22\textwidth]{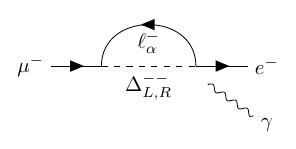}
\includegraphics[width=0.22\textwidth]{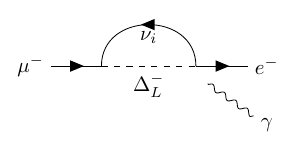}
\caption{Representative Feynman diagrams for $\mu \to e\gamma$. The photon can be radiated from any charged external leg or internal propagator.
}
\label{fig:mu2e}
\end{figure}

\begin{figure}[!htbp]
\centering
\includegraphics[width=0.22\textwidth]{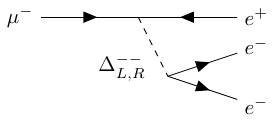}
\caption{Representative Feynman diagram for $\mu \to e e e$. 
}
\label{fig:mueee}
\end{figure}

\begin{figure}[!htbp]
\centering
\includegraphics[width=0.22\textwidth]{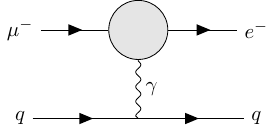}
\includegraphics[width=0.22\textwidth]{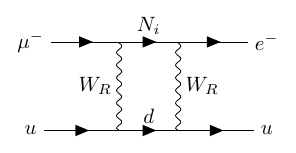}
\caption{Representative Feynman diagrams for $\mu \to e$ conversion in nuclei. The effective $\mu e\gamma^*$ vertex (shaded circle) incorporates contributions from the diagram shown in Fig.~\ref{fig:mu2e}.}
\label{fig:mueconv}
\end{figure}

The CLFV processes $\mu \to e \gamma$, $\mu \to e e e$ and $\mu \to e$ conversion are depicted in Figs.~\ref{fig:mu2e},~\ref{fig:mueee} and~\ref{fig:mueconv}, respectively. 
The decay branching ratios of the former two are given by~\cite{Cirigliano:2004mv,Barry:2013xxa}
\begin{align}
\label{eq:br_muea}
\mathrm{BR}(\mu \to e\gamma) &= \frac{3\alpha_{\mathrm{em}}}{2\pi} \left( |G_L^\gamma|^2 + |G_R^\gamma|^2 \right)\,,\\
{\rm Br}(\mu \to eee)
\label{eq:br_mu3e}
&= \frac{1}{2}\, |h_{e\mu} h_{ee}^{*}|^2 \left(
\frac{m_{W_L}^4}{m_{\Delta_R^{--}}^4}
+ \frac{m_{W_L}^4}{m_{\Delta_L^{--}}^4}
\right)\,.
\end{align}
Here, $\alpha_{\mathrm{em}} = e^2/(4\pi)$ is the electromagnetic coupling constant, the form factors $G_L^{\gamma}$ and $G_R^{\gamma}$ are expressed as
\begin{align}
G_L^{\gamma} &= \sum_i V_{\mu i} V_{ei}^* \left( \frac{m_{W_L}^2}{m_{W_R}^2} G_1^{\gamma}(y_i) + \frac{2y_i}{3} \frac{m_{W_L}^2}{m^2_{\Delta_{R}^{--}}} \right)\,, \\
G_R^{\gamma} &= \sum_i V_{\mu i} V_{ei}^* y_i \left( \frac{2}{3} \frac{m_{W_L}^2}{m^2_{\Delta_{L}^{--}}} + \frac{1}{12} \frac{m_{W_L}^2}{m^2_{\Delta_{L}^{-}}} \right)\,,
\end{align}
with $x_i \equiv (m_{N_i}/m_{W_L})^2$, $y_i \equiv (m_{N_i}/m_{W_R})^2$.
The loop function $G_1^{\gamma}(y)$ is given as
\begin{align}
G_1^{\gamma}(y) &= -\frac{2y^3 + 5y^2 - y}{4(1 - y)^3} - \frac{3y^3}{2(1 - y)^4} \ln y\,.
\end{align}
The effective coupling is defined as 
\begin{align}
h_{\alpha\beta} &= \sum_{i} V_{\alpha i} V_{\beta i} \left( \frac{m_{N_i}}{m_{W_R}} \right), \qquad \alpha,\beta = e, \mu\,.
\end{align}

If the heavy particles $W_R$, $\Delta_{L,R}^{--}$ and $\Delta_L^-$ are degenerate in mass, the $\mu N \to e N$ conversion branching ratio is approximated by~\cite{Cirigliano:2004mv}
\begin{align}
\label{eq:br_mu2e}
R_N(\mu \to e )
&\simeq X_N \times 10^{-7}\, |g_{e\mu}|^2
\left( \frac{1\,\mathrm{TeV}}{m_{\Delta_R^{--}}} \right)^4 \notag \\
&\quad \times \alpha_{\mathrm{em}}
\left( \ln \frac{m_{\Delta_R^{--}}^2}{m_\mu^2} \right)^2\,,
\end{align}
where the factor $X_N$ depends on the nucleus, with $X_{(\text{Al},\text{Ti},\text{Au})}\simeq (0.8,\,1.3,\,1.6)$, and $m_\mu$ denotes the mass of muon. 
The effective coupling is defined as
\begin{align}
g_{e\mu} &= \sum_i V_{ei}^* V_{\mu i} \left( \frac{m_{N_i}}{m_{W_R}} \right)^2\;.
\end{align}
Notice that, in addition to the tree-level diagram shown in Fig.~\ref{fig:mueee}, there exist additional one-loop contribution arising from the effective $\mu e\gamma^*$ vertex, which is much suppressed~\cite{Cirigliano:2004mv}.

\newpage

\bibliographystyle{apsrev4-1}
\bibliography{reference.bib}% Produces the bibliography via BibTeX.

\end{document}